\documentclass[smallcondensed]{svjour3}    
\usepackage{enumitem}
\smartqed  

\usepackage{graphicx}
\begin{document}

\title{Exchange orbits - an interesting case of co-orbital motion}

\titlerunning{Exchange orbits} 

\author{Barbara Funk \and
       Rudolf Dvorak         \and
        Richard Schwarz
}


\institute{ \at
              T\"urkenschanzstarsse 17, 1180 Vienna \\
              Tel.: +43-1-427751840\\
              \email{b.funk@univie.ac.at}\\
              \email{rudolf.dvorak@univie.ac.at}\\
              \email{richard.schwarz@univie.ac.at} }        
\maketitle

\begin{abstract}

In this investigation we treat a special configuration of two celestial bodies
in 1:1 mean motion resonance namely the so-called exchange orbits. There exist
-- at least -- theoretically -- two different types: the exchange-a orbits and
the exchange-e orbits. The first one is the following: two 
celestial bodies are in orbit around a central body with almost the same
semi-major axes on circular orbits. Because of the relatively small
differences in semi-major axes they meet from time to time and exchange their
semi-major axes. The inner one then moves outside the other planet and vice
versa. The second configuration one is the following: two planets are moving on
nearly the same orbit with respect to the semi-major axes, one on a 
circular orbit and the other one on an eccentric one. During their dynamical
evolution they change the characteristics of the orbit, the circular one becomes
an elliptic one whereas the elliptic one changes its shape to a circle. This
'game' repeats periodically. In this new study we extend the numerical
computations for both of these exchange orbits to the three dimensional case
and in another extension treat also the problem when these orbits are
perturbed from a fourth body. Our results in form of graphs show quite well
that for a large variety of initial conditions both configurations are stable
and stay in this exchange orbits.
  

\end{abstract}

\section{Introduction}
\label{intro}

Speaking of co-orbital motion we mean that two celestial bodies are in a 1:1 mean motion resonance (MMR)
with respect to each other in orbiting a central body. We know that this is a common phenomenon in
our Solar System especially for asteroids which are the so-called Trojans;
such an asteroid may move close to one of the stable
equilibrium points $L_4$ or $L_5$. The Lagrangian equilibrium points are the five
stationary solutions of the restricted three body problem (=RTBP). 
The points $L_1$, $L_2$, and $L_3$ lie on a straight line connecting the two
primary bodies and are points of unstable equilibrium. The Lagrangian points 
$L_4$ or $L_5$ are in the isosceles configuration and are only stable for a
  mass parameter $\mu=m_2/(m_1+m_2)$, where $m_1$ and $m_2$ are masses of
  the primaries, with $m_2$ smaller than $m_1$ ($\mu < 1/25$). In the Solar System there
  exist three planets hosting Trojan asteroids around $L_4$/$L_5$ in the 1:1 mean motion
  resonance namely Mars, Jupiter and Neptune. Concerning the Trojan motion there has been undertaken a big number of investigations 
in theoretical respects as 
well as in numerical studies for our Solar System but also for extrasolar
planets.\\
This work is mainly dedicated to the following investigation: which dynamical configurations are stable for planets in a special kind of co-orbital motion - the so-called exchange orbits. Here we understand two types of the exchange of the orbital elements of two planets, on one hand two celestial bodies in orbit around a central star with almost the same
semi-major axes on circular orbits (which we call exchange-a (xch-a) orbits) and on the other 
hand two planets with different eccentricities but the same semi-major axes (which we call exchange-e (xch-e) orbits). In both configurations a periodical change happens with respect to the semi-major axis (xch-a orbits) respectively the shape of their orbits, their ellipticity (e-orbit). In particular Laughlin \& Chambers, 2002 give an introduction to all three kinds of co-orbital motion (Trojan, {\bf Exchange-a} and {\bf Exchange-e} motion). Numerical studies concerning the {\bf Exchange-e} orbits were done by Roth, 2009 and Nauenberg, 2002. Nauenberg, 2002 discussed the possibilities to detect such orbits in extrasolar planetary systems. He showed that with the aid of RV measurements {\bf Exchange-e} orbits could be distinguishable from the case of a single planet and that they are long-term stable. Funk et al., 2011 focused on the {\bf Exchange-e} configuration and determined the stable regions in dependence on the mean anomaly and the eccentricity of the planets.\\
While for the {\bf Exchange-e} configuration only theoretical studies exist we have a real example for the {\bf Exchange-a} configuration - namely the two Saturnian moons Janus and Epimetheus. They are in a 1:1 orbital resonance and perform a horseshoe orbit. There exist already some theoretical studies concerning the topic of {\bf Exchange-a} orbits. For example Dermott \& Murray, 1981, Spirig \& Waldvogel, 1985, Waldvogel \& Spirig, 1988 and Auner, 2001 \cite{auner} studied the horseshoe motion by perturbation theory and numerical experiments. Yoder et al., 1983, 1989 developed an analytic approximation for the {\bf Exchange-a} motion. Llibre \& Oll\'e, 2001 proved the existence of stable planar horseshoe periodic orbits for the Saturn-Janus mass parameter in the restricted circular three-body problem. While first studies used the restricted three body problem, Cors \& Hall, 2003 \cite{cors} and Bengochea \& Pi\~na, 2009 \cite{pina} took also into account the general three body model. Barrab\'es \& Mikkola, 2005 found families of periodic orbits in the planar model and also took into account the spatial case. Dvorak, 2006 \cite{dvorak} did numerical experiments for different masses of the two planets involved and different initial separation of the semi-major axis. It turned out that for stable exchange orbits the sum of the mass of the two planets can only slightly exceed the one of Saturn.

The organization of this article is as follows:
We will study in detail the xch-a orbits where only few theoretical work exists and will also shortly show what has been found up to know concerning the 
stability character of these kind of orbits when the two planets have
different masses. In our extension of this work we will concentrate on
inclined orbits and on perturbations of an additional massive body 
inside or outside the orbits of the two celestial bodies in a 1:1 MMR.

We then will briefly report the known results about the xch-e orbits but will extend the investigation in two directions: we will show how the inclinations 
change the orbital characteristic and in addition we will numerically test
their stability when other (more massive) planets will perturb their orbits.\\
In the final discussion we see how our results are interesting for extrasolar planets where up to now so many different 'architectures' of such systems were found.

\section{EXCHANGE-A ORBITS} 

\begin{figure}
\centering
\includegraphics[width=7cm]{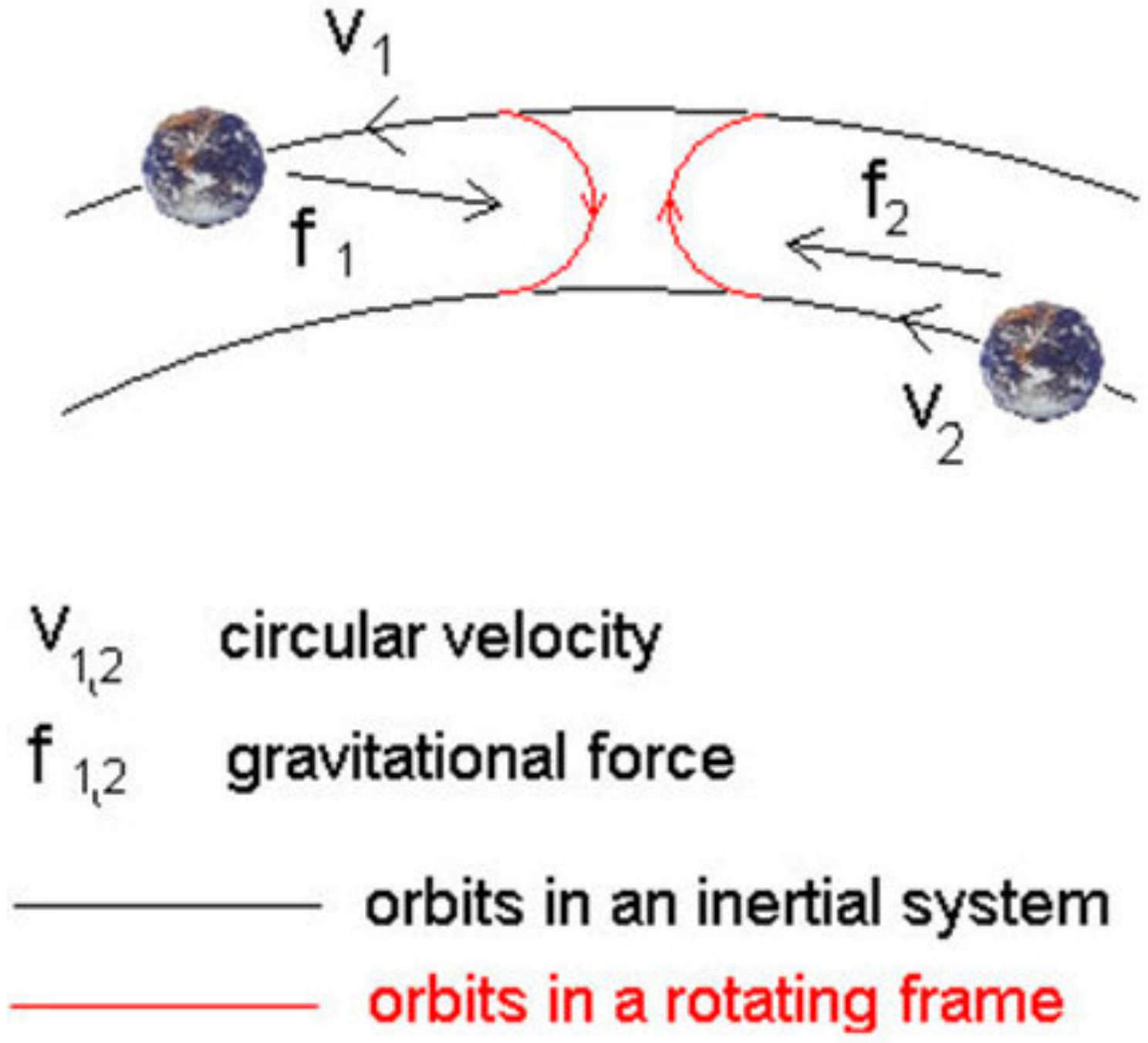}
\caption{The graph presents the interaction of two planets in an xch-a configuration.}
\label{fig2}
\end{figure}

One can define this configuration as follows: two planets move on nearly circular
  orbits with almost the same semi-major axis around a much more massive host
  (star). Because of the 3rd Keplerian law the one moving on the inner orbit is
  faster and approaches the outer one (planet) from behind. Before they meet,
  the inner planet is shifted to the orbit of the outer (Fig. \ref{fig2}) and the outer planet 
moves inward (orbit with a smaller semi-major axis (a)),
  that means that they have changed their orbits. The xch-a configuration (e.g. Spirig \& Waldvogel, 1985, Auner, 2001) can be found in our Solar System, the two moons of Saturn,
  Janus and Epimetheus show such a motion.

\subsection{Method and numerical setup}
\label{model-a}
With the aid of numerical integrations (Lie-method with adaptive step size, e.g. Hanslmeier and Dvorak, 1984, Lichtenegger, 1984, Delva, 1985) the equations of motion in a pure
gravitational problem (3 respectively 4-body problem) using point masses for
the central star and the planets involved we derived our results.
The dynamical models and the initial conditions were the following ones:

\begin{itemize}
\item {\bf The 3-body problem:}  Sun and two planets P1 and P2 in xch-a orbits;

\begin{itemize}
\item we varied the initial separation in semi-major axes of the two planets
from $0 < \delta a < 0.025$ in steps of $\delta a = 0.001$.
\item we changed the mass ratio between the planets $m_{P2}/m_{P1}$ between
1 and 100, where $m_{P1}$ has always 1 $m_{Earth}$ and $m_{P2}$ was changed from 1 to 100 $m_{Earth}$.
\item we changed the initial inclination for the orbital planes of P1 and P2 between
$0^{\circ}<i<56^{\circ}$
\end{itemize}

\item {\bf The 4-body problem:} Sun and two Earth-like planets and:

\begin{itemize}
\item a perturbing Jupiter inside the orbit of P1 and P2 (for P1 and P2 in the
same orbit with an inclined P3 and as second case P1 and P3 in one orbital plane and an
inclined P2)
\item a perturbing Jupiter outside the orbit of P1 and P2 with inclinations like
mentioned before in ii).
\end{itemize}
\end{itemize}

The osculating elements were chosen such that all orbits are initially
circular ($e=0$), for the mean anomalies for the three planets we have taken
$M_{P1}=0^{\circ}$, $M_{P2}=180^{\circ}$ and  $M_{P3}=90^{\circ}$; $\omega$
and $\Omega$ were set to the same values for all 2 (3) planets. The normal
duration of the integration was set to $10^6$ years for the unperturbed
problem and $10^5$ years for the perturbed problem. Note that the 'nominal'
semi-major axes for the planets in exchange orbits was always set to a=1AU and
we used 'astrocentric' coordinates (elements)
 
\subsection{Orbits for unperturbed cases: a three-body problem}

\begin{figure}
\centering
\includegraphics[width=4.1cm,angle=270]{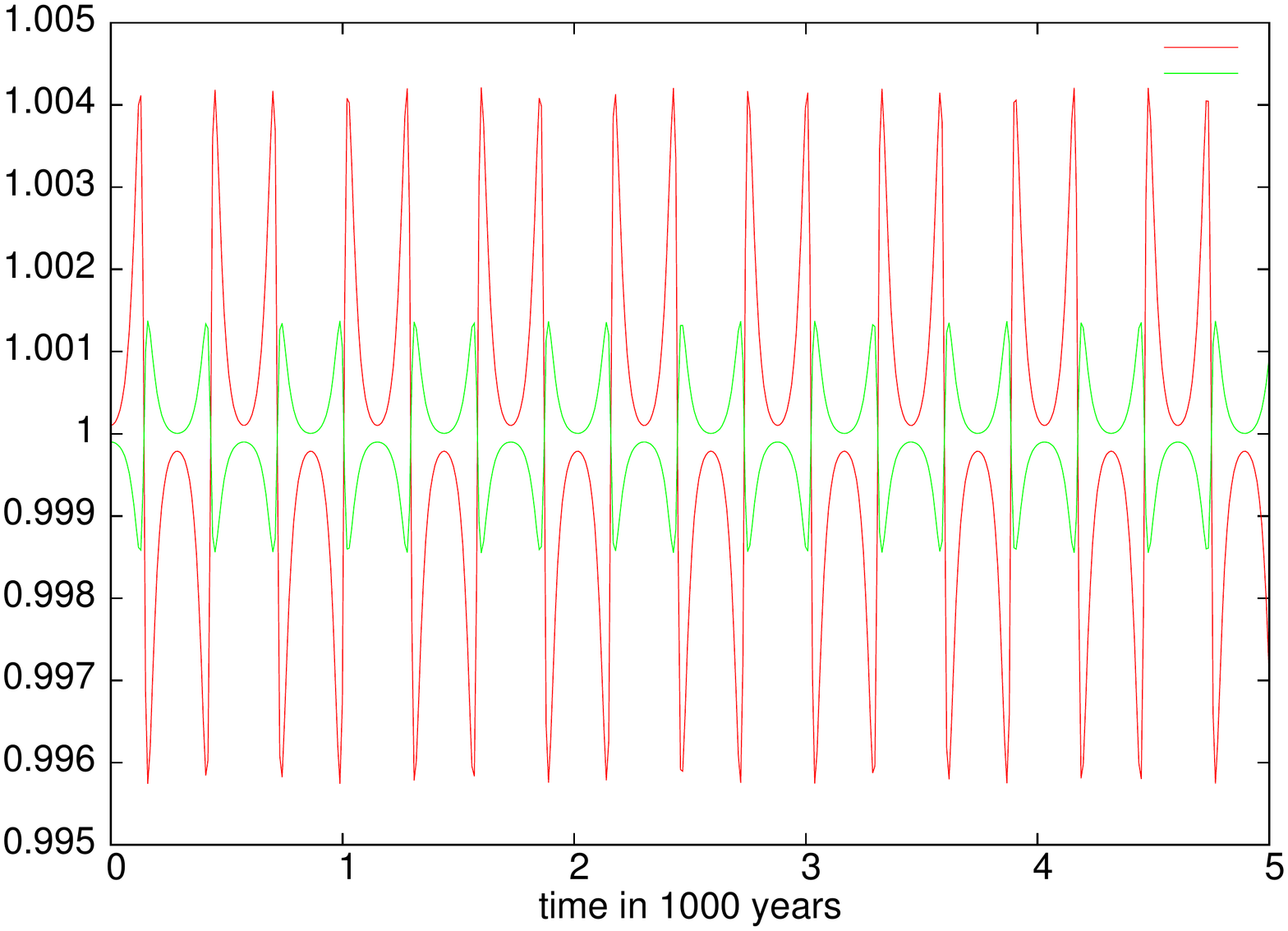} 
\includegraphics[width=4.1cm,angle=270]{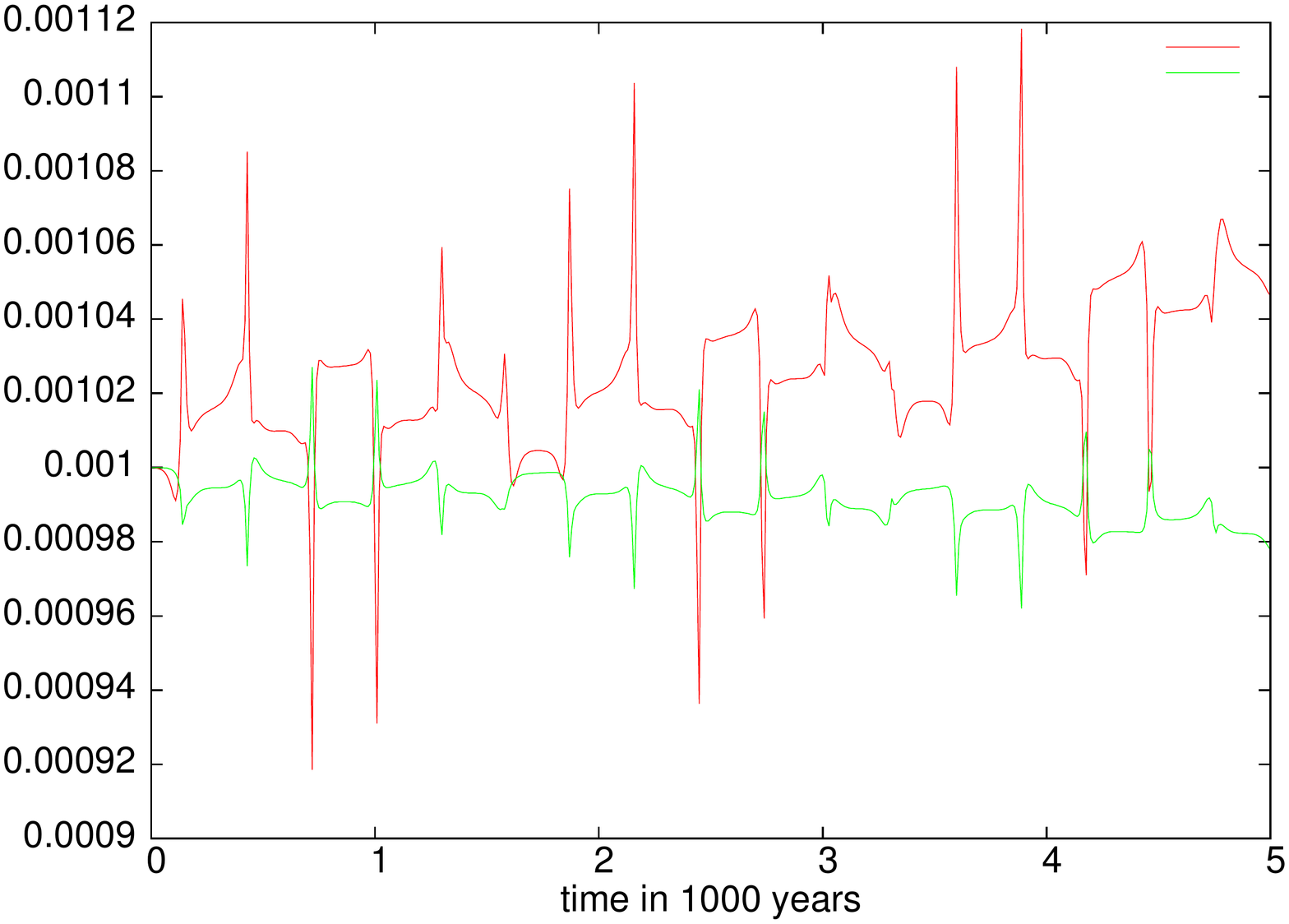}
\caption{Dynamical evolution of xch-a orbits for an initial 
inclination  $i=9^{\circ}$ for a planet pair with P2 (green)
  18 times more massive than P1 (earth-sized, red). Semi-major axes (left graph)
respectively eccentricity (right graph) versus time.}
\label{m3-i9}
\end{figure}

We start with some examples of xch-a orbits for different models.
In a first integration we fixed the masses $m_{P2}= 18 \cdot m_{P1}$ and for the
initial inclination we set  $i=9^{\circ}$. In Fig.\ref{m3-i9} we observe
the regular exchange of the two orbits in the sense described above: the red
curve (larger amplitudes) shows the semi-major axis (left graph) of P1, the green curve 
the one of the heavier planet P2. At the moment of exchange the differences in
semi-major axes achieve the largest values, because at the instant of close
encounter the additional acceleration causes a sudden increase in velocity; 
this is also visible from the spikes in the eccentricity (right
graph). Although here we show only the regular time evolution for $5 \cdot
10^3$ years the orbits stay in such a stable configuration at least for millions of years as
have shown our own test computations.

\begin{figure}
\centering
\includegraphics[width=4.1cm,angle=270]{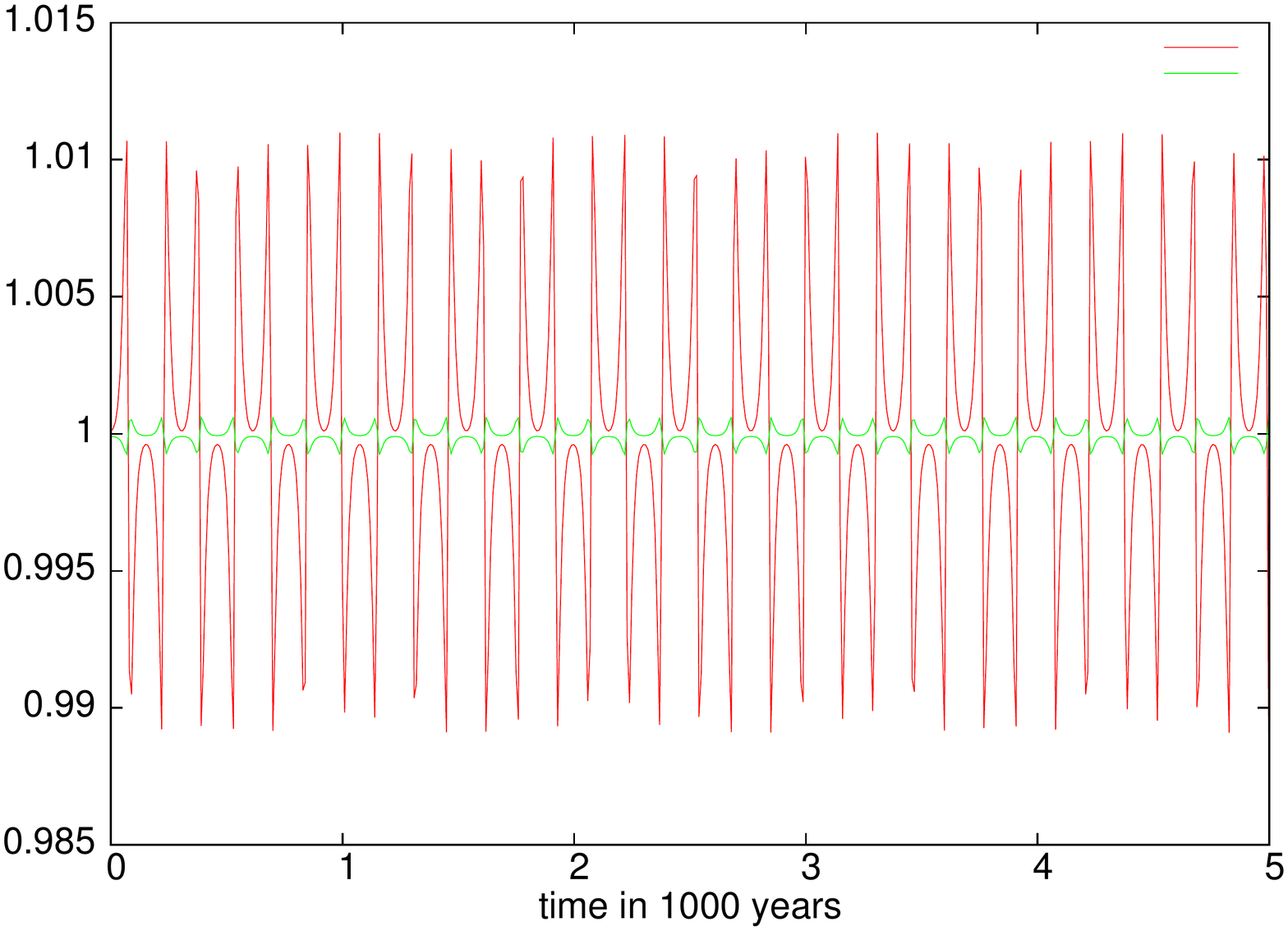}
\includegraphics[width=4.1cm,angle=270]{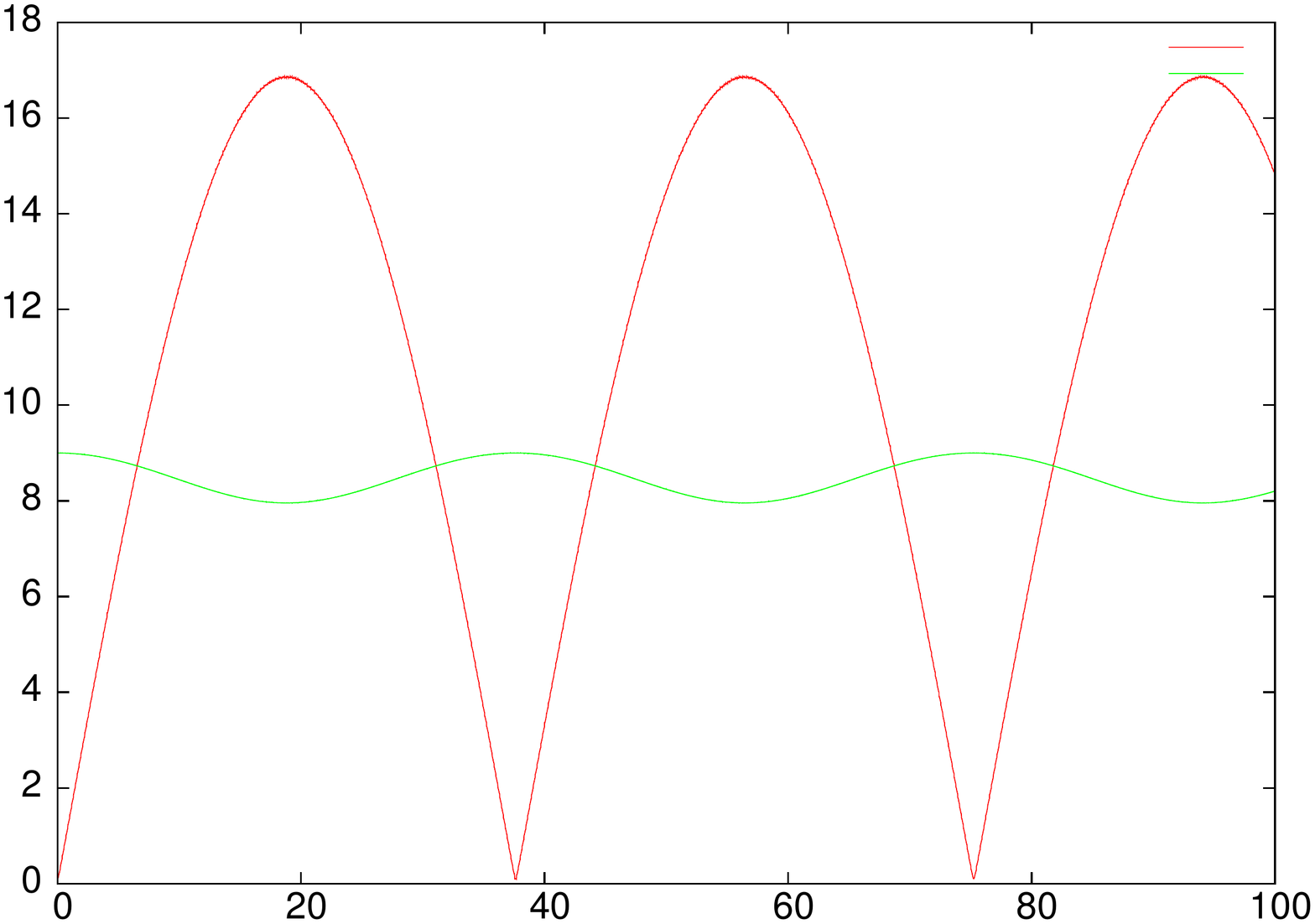}
\caption{Dynamical evolution of an xch-a orbit 
for a planet pair 
Earth (P1, red) -- Saturn (P2, green) for an inclination 
of $i=9^{\circ}$; semi-major axes (left graph), respectively inclination
(right graph) versus time.}
\label{m16-i9}
\end{figure}

In the next examples, again for  $i=9^{\circ}$, we changed the two masses such
that $m_{P2} \sim 100 \cdot m_{P1}$ which is a planet pair Earth -- Saturn. One can
observe in Fig. \ref{m16-i9} (left graph) how often the two planets meet and change their
positions with respect to the semi-major axis compared to the case 
$m_{P2}= 18 \cdot m_{p1}$ (Fig.\ref{m3-i9}, left graph). Surprising is the
long period (note that the time scale
for the two graphs is different!) of the respective change of
the inclinations (Fig. \ref{m16-i9} right graph) because one would expect
a certain coupling between eccentricities and inclinations. In fact here the
eccentricity are so small that they do not count for the stability of the
angular momentum. One can very well see the small
variations of the inclination of the
'heavy Saturn' compared to the 'Earth'.

\begin{figure}
\centering
\includegraphics[width=4.1cm,angle=270]{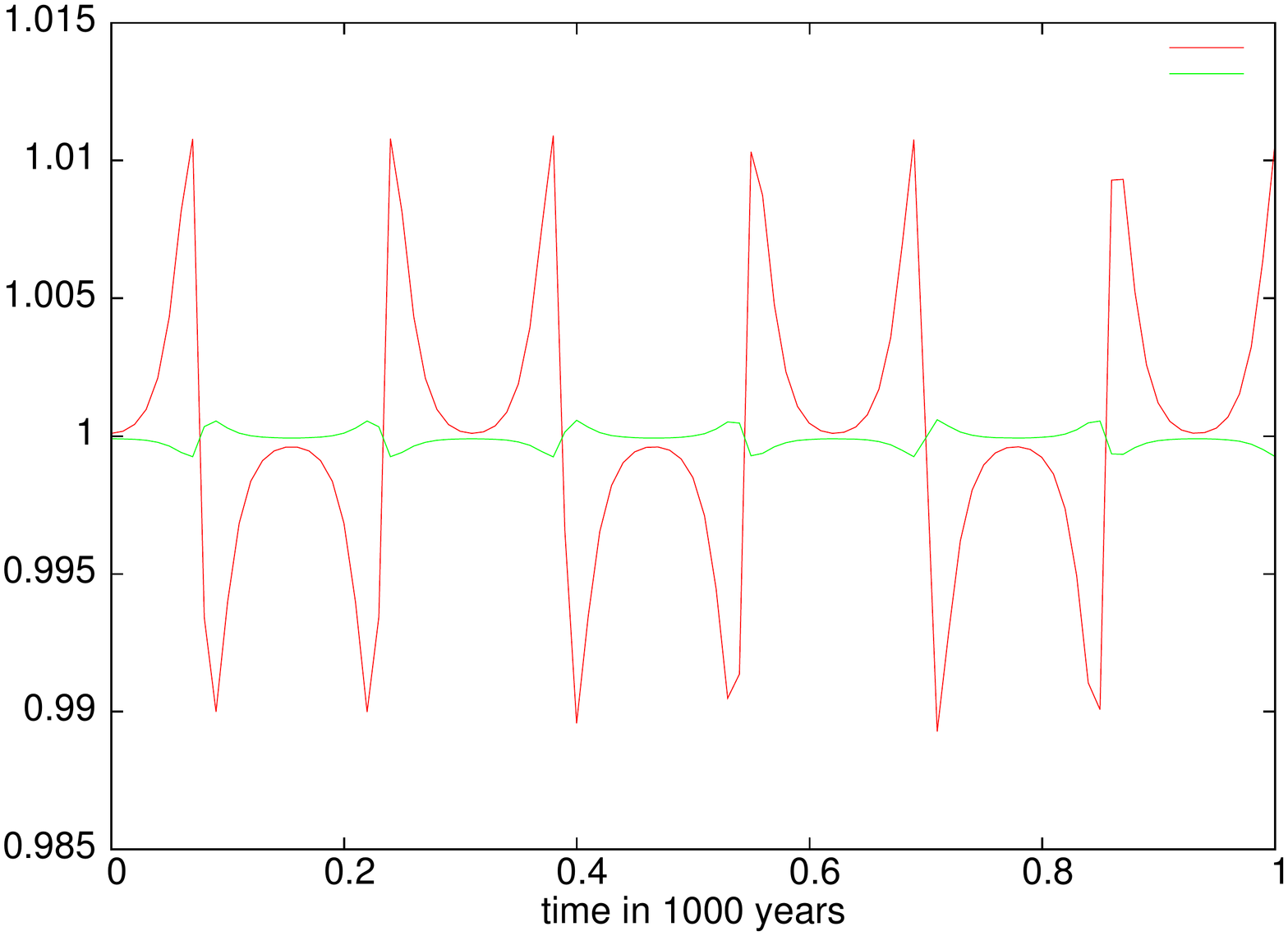}
\includegraphics[width=4.1cm,angle=270]{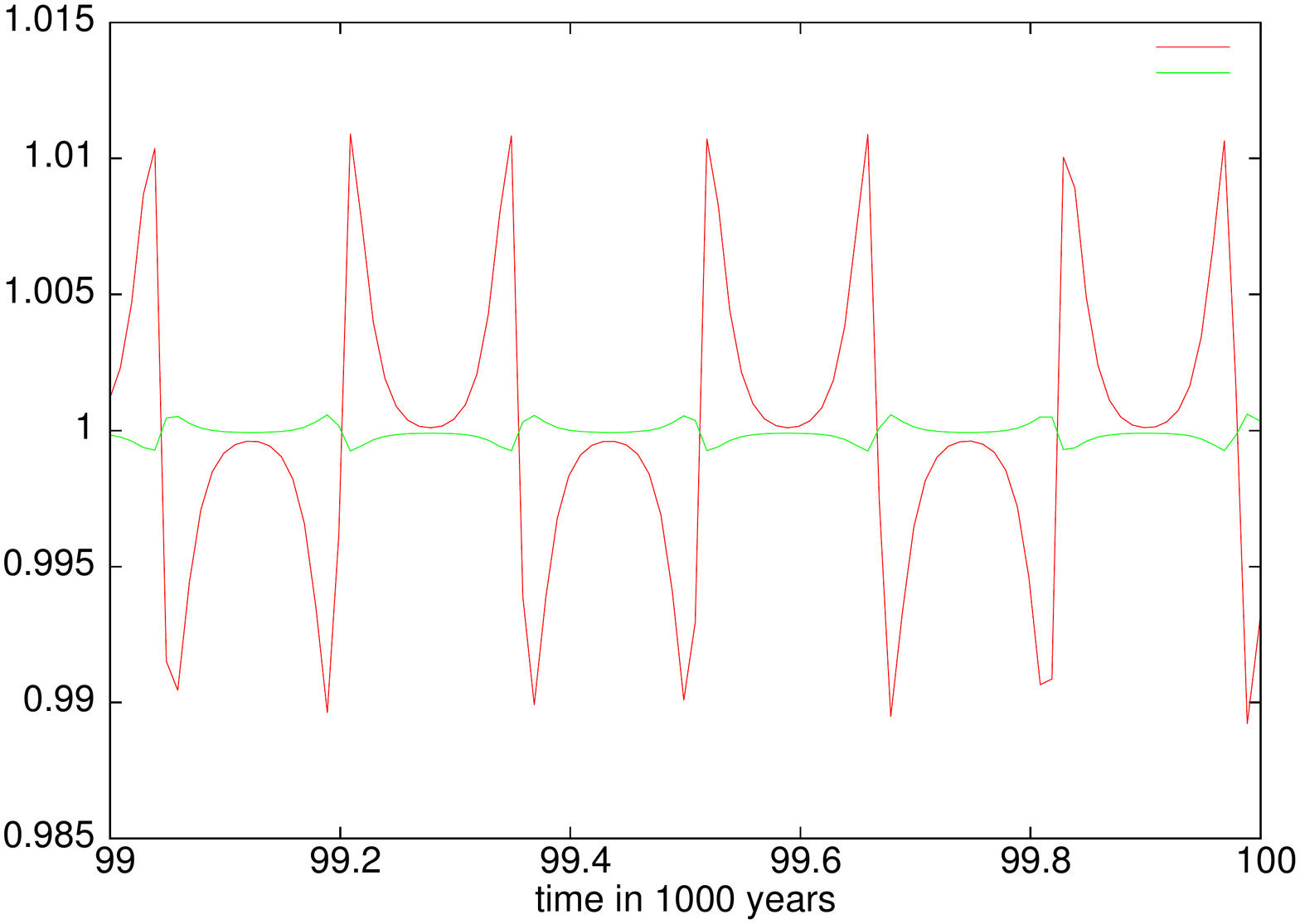}
\caption{Dynamical evolution of xch-a orbit 
for a planet pair 
Earth (P1, red) -- Saturn (P2, green) for an inclination 
of $i=17^{\circ}$. Semi-major axes for the first 1000 years (left graph) and for the last 1000 years (right graph).}
\label{m16-i17a}
\end{figure}

The stability of the orbits over long time even with large inclinations
(in this example we set $i=17^{\circ}$ for a Earth-Saturn pair) can be seen in Fig.\ref{m16-i17a}
where we compare the dynamical behavior of
the semi-major axes for the first $10^3$ years (left graph) with the last
$10^3$ years in a $10^5$ years integration of the orbits. The signal is
exactly identical with respect to the period and the amplitudes.\\

\begin{figure}
\centering
\includegraphics[width=4.1cm,angle=270]{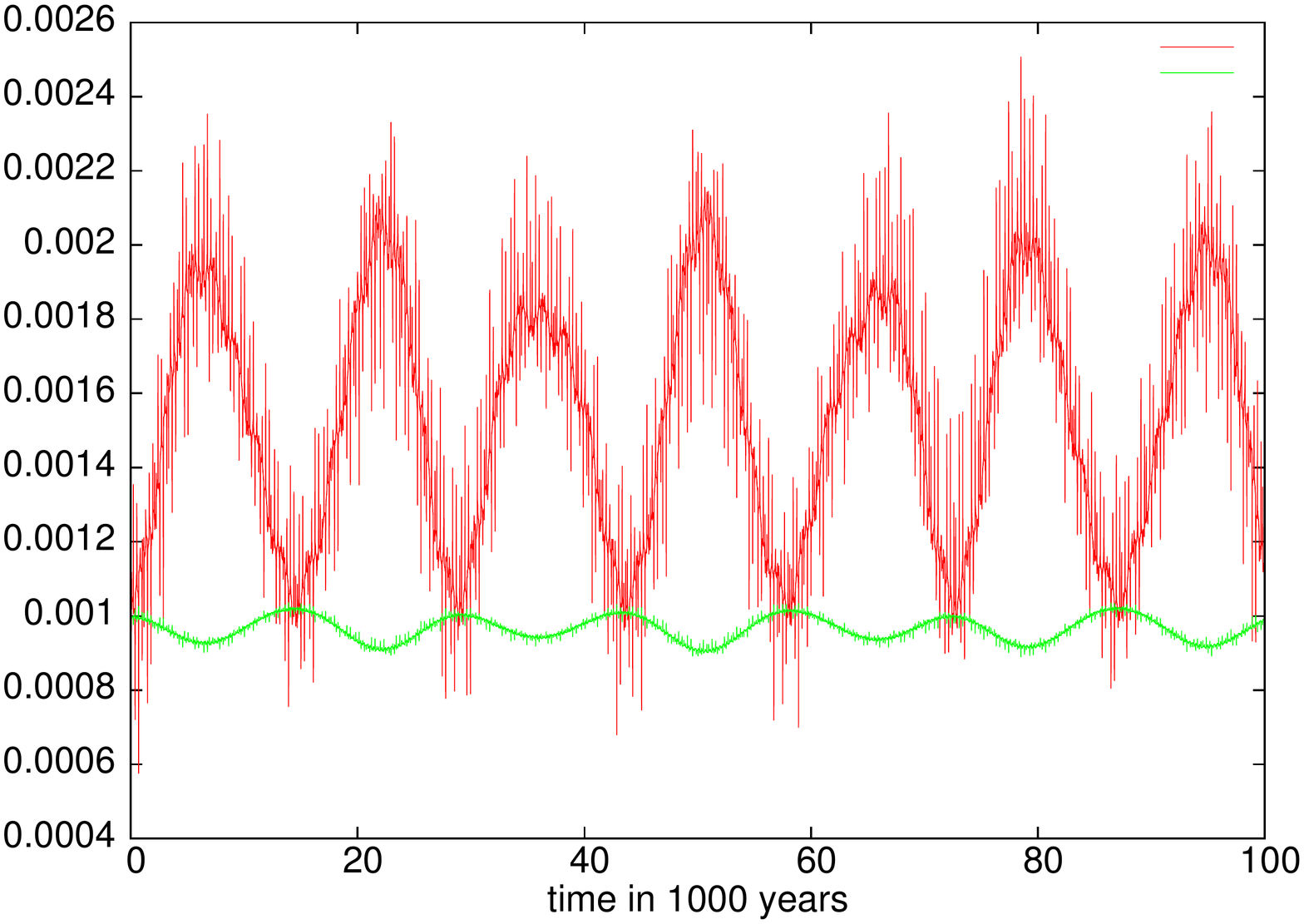}
\includegraphics[width=4.1cm,angle=270]{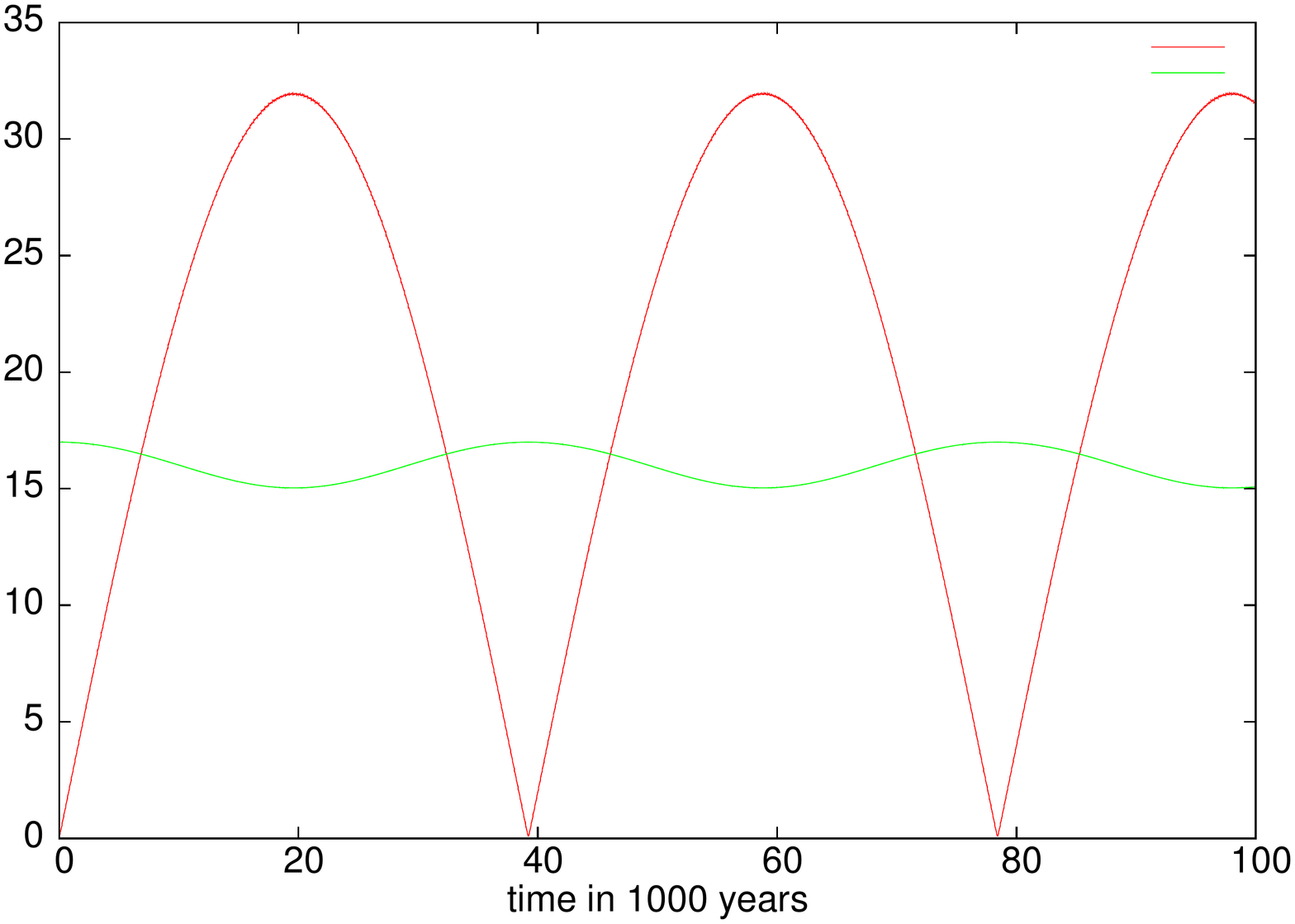}
\caption{Dynamical evolution of xch-a orbits 
for a planet pair 
Earth (P1, red) -- Saturn (P2, green) for an inclination 
of $i=17^{\circ}$: eccentricity (left graph) and inclination (right graph).}
\label{m16-i17}
\end{figure}

Finally again for an integration of $10^5$ years and the same model as before we 
plotted the eccentricities of both planets involved (Fig.\ref{m16-i17}, left graph) 
where a period of some 
$14 \cdot 10^3$ years is well visible for Saturn (green curve with small
amplitudes) and the Earth (red curve with large amplitudes).
The evolution of the inclination (Fig. \ref{m16-i17a}, right graph) is very similar in period with
the one in Fig.\ref{m16-i9} (right graph); it shows that the inclinations change
quite large for the smaller planet and reaches values up to twice the
inclination of the larger planet. Comparing the periods in detail for these two
different inclinations (i = $9^{\circ}$ and $17^{\circ}$) only very small
differences are visible. We can conclude that the orbits not only exchange the
semi-major axis and the eccentricities but also the inclinations.

\begin{figure}
\centering
\includegraphics[width=6.2cm,angle=270]{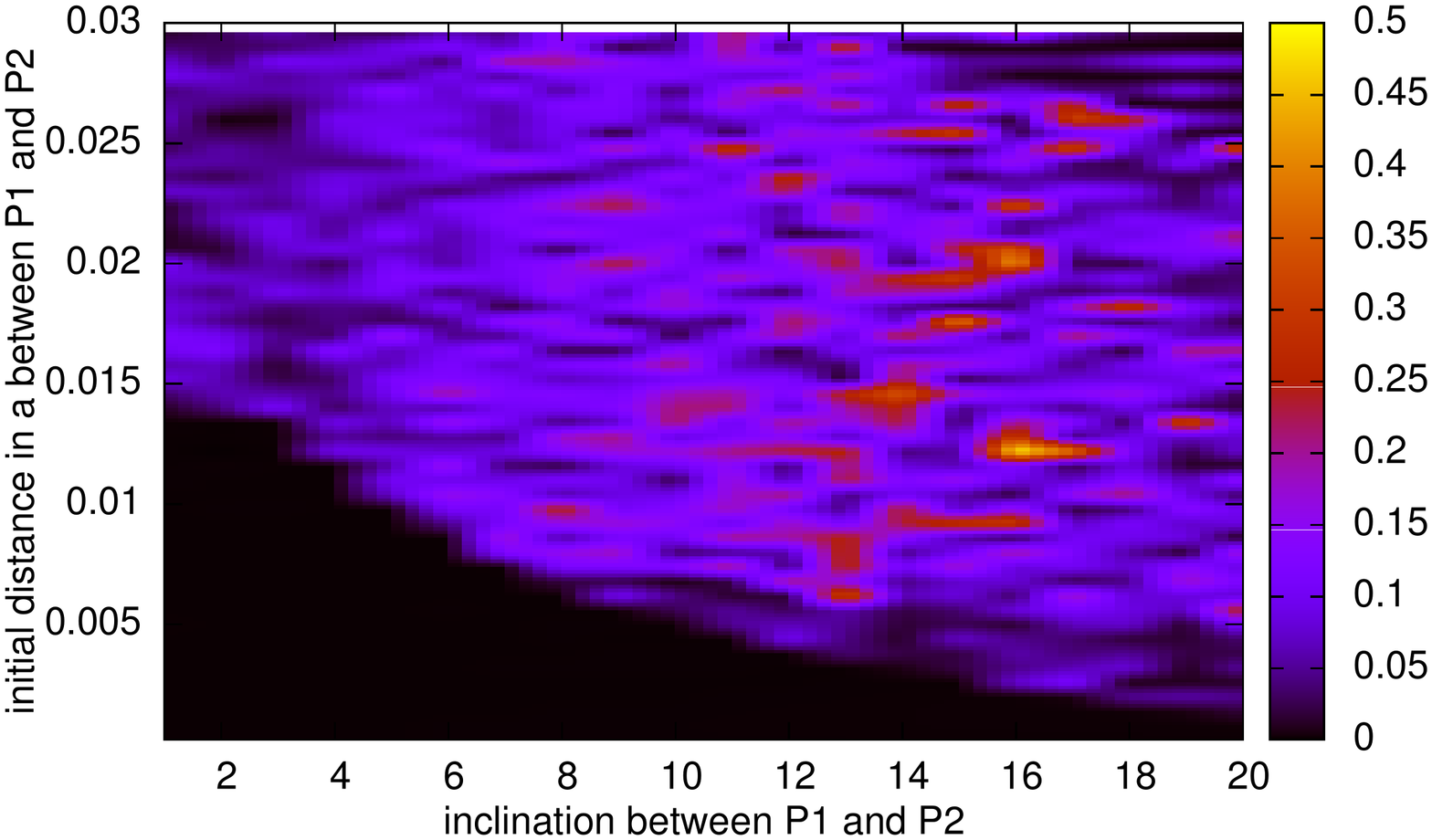}
\caption{The shrinking of the stable region of the xch-a orbits:
x-axes shows the initial inclination of the two orbits, the y-axes
shows the difference in semi-major axes. The color code gives the maximum
eccentricity, where black corresponds to stable motion in the sense that they
are still in the same exchange orbits for the whole integration time; for more
see in the text} 
\label{inclination}
\end{figure}

To get an overall picture of the dependence of the stable region on the separation
in semi-major axes and inclination of P1 and P2 we
show the results of our computations in the three body problem. We started in 
different separations of the semi-major axes (see section \ref{model-a}) and for different inclinations
($0^{\circ} < i < 20^{\circ}$). 
From the respective graph (Fig.\ref{inclination}) one can
see that up to $i=4^{\circ}$ this separation leads to stable xch-a orbits 
up to 0.03 AU ($\delta$a = $\pm$ 0.015, note that the separation
is counted from a=1 AU in both directions). Then a steady -- almost linear --
decrease for the largeness of the stable region is visible and from
$i=20^{\circ}$ on the two planets are not any more in stable xch-a
orbits.\\

\begin{figure}
\centering
\includegraphics[width=4.2cm,angle=270]{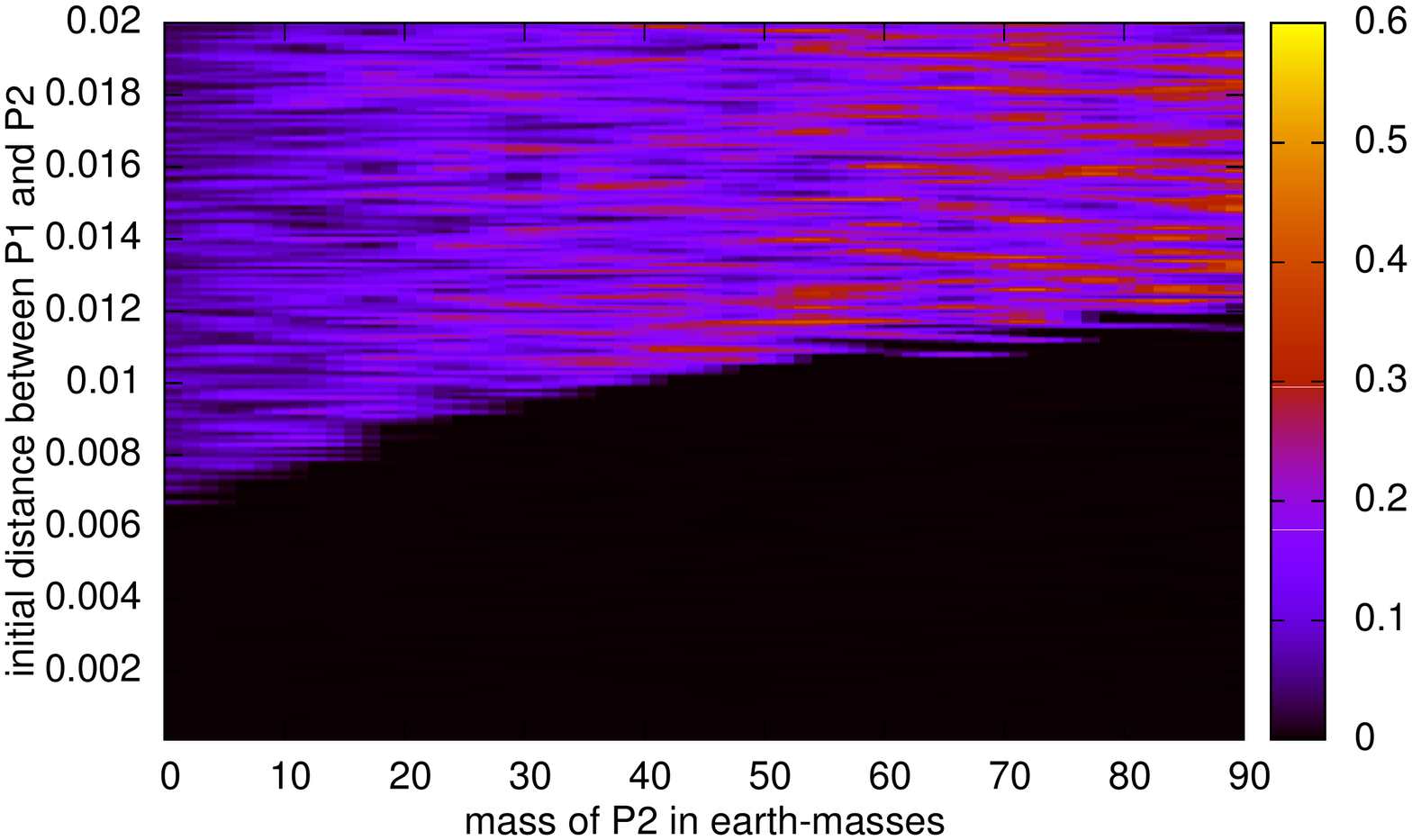}
\includegraphics[width=4.2cm,angle=270]{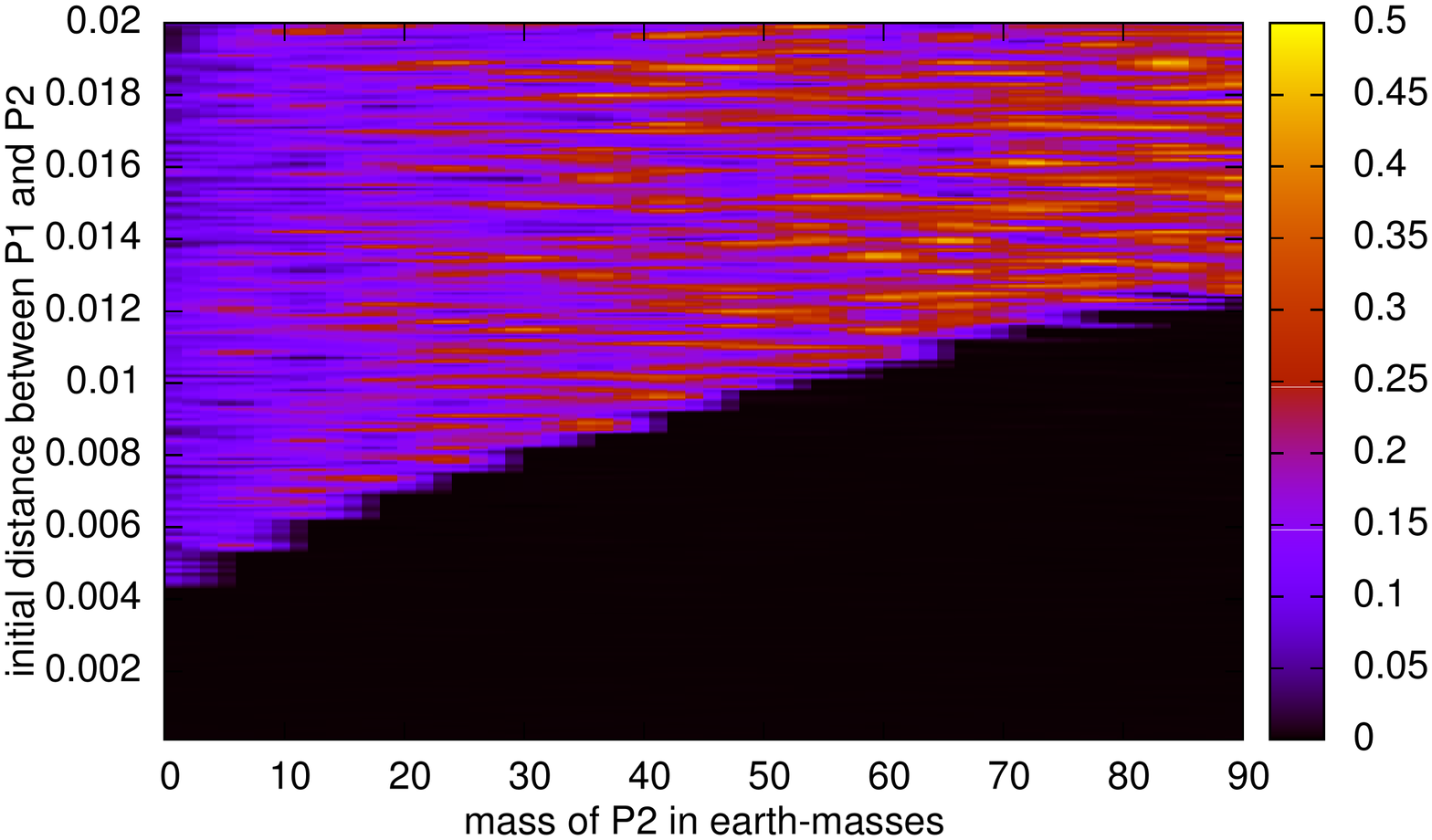}
\includegraphics[width=4.2cm,angle=270]{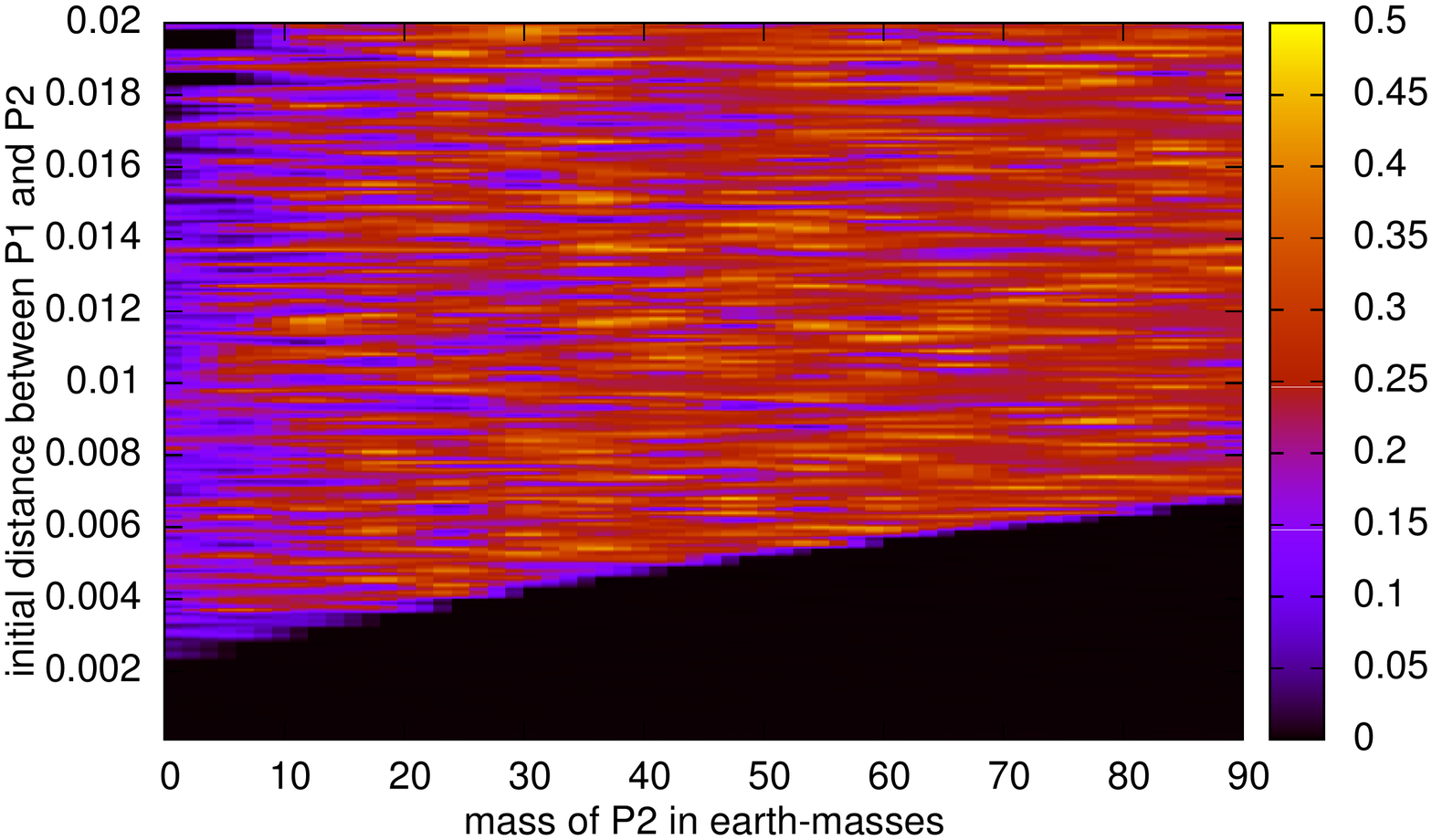}
\includegraphics[width=4.2cm,angle=270]{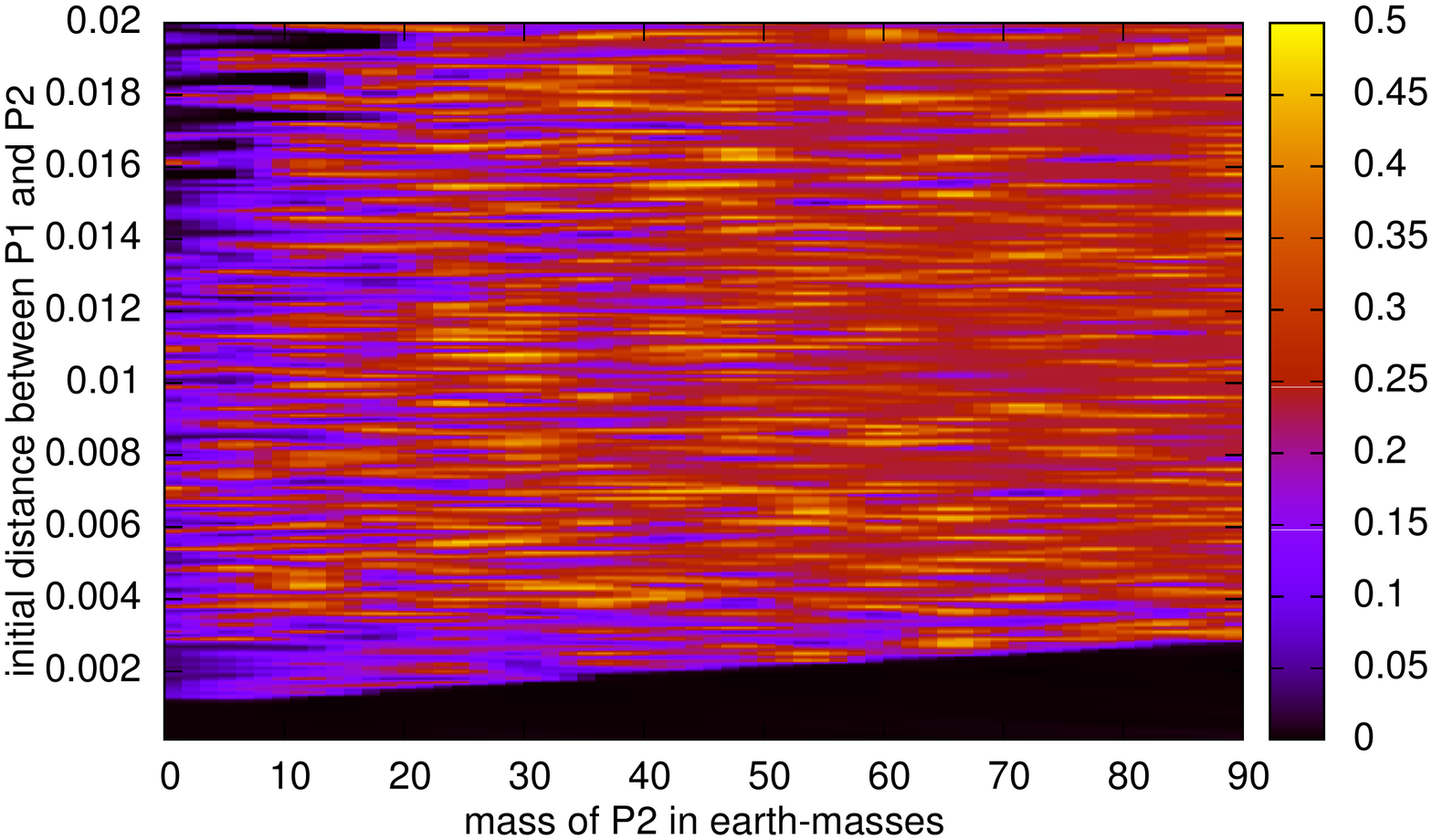}
\caption{The shrinking of the stable region of the xch-a orbits: x-axes
  mass of planet P2, y-axes $\delta a$ for different inclinations of the two
  planets (top, left: $i = 1^{\circ}$; top, right: $i = 6^{\circ}$; lower,
  left: $i = 11^{\circ}$; lower, right: $i = 16^{\circ}$; color code like in
  (Fig.\ref{inclination}}
\label{masses}
\end{figure}

We have also undertaken a test of stable orbits depending on one hand on the
masses of the two planet and on the other hand we also changed the initial inclination $i$. In Fig. \ref{masses} it is quite well
visible how this region shrinks with larger inclinations, but stays
relatively large for inclinations  $i < 8^{\circ}$ (not shown) but then shrinks and 
disappears completely for $i > 20^{\circ}$. It is quite interesting that with
a larger planet P2 the stable region increases.

\subsection{Perturbed xch-a orbits: a four body problem}
Because we were interested how stable this configuration can be under the perturbation of an additional large planet P3 (Jupiter-sized) we have undertaken the following 4 different runs:
\begin{itemize}
\item {\bf (a)}: A perturbing planet P3 from inside the orbits of P1 and P2 where we changed the  inclination of P1 up to $i=56^{\circ}$; P3 shared the orbital plane with P2.
\item {\bf (b)}: A perturbing planet P3 from inside the orbits of P1 and P2 where we changed the inclination of P3 up to $i=56^{\circ}$; P1 and P2 are in the same plane.
\item {\bf (c)}: A perturbing planet P3 from outside the orbits of P1 and P2 where we changed the inclination of P2 up to $i=56^{\circ}$; P3 shared the orbital plane with P1.
\item {\bf (d)}: A perturbing planet P3 from outside  the orbits of P1 and P2 where we changed the inclination of  P3 up to $i=56^{\circ}$; P1 and P2 are in the same plane.
\end{itemize}

\subsubsection{Results (a) and (b): an inner Jupiter P3 perturbs P1-P2}
\begin{enumerate}[label={(\alph*)}]
\item We started with computations in the four-body problem with a perturbing inner
Jupiter,
where we varied the distance in steps of $\Delta a_{P3}=0.01$ for $0.4$ AU $<$
$a_{P3} < 0.8$ AU  for eight different inclinations for P1 with respect to the
orbit of P1-P2 for $0^{\circ} < i < 56^{\circ}$ ($\Delta$i = $8^{\circ}$). For every of the 8
different inclinations we also varied the separation from 0.0025 AU $<$ $\delta a_{P1}$ $<$ 0.1 AU\footnote{Note that for the semi-major axis of P1 we have taken 1~AU~-~$\delta a_{P1}$ and for P2 1~AU~+~$\delta a_{P1}$} with a step of $\Delta \delta a_{P1} = 0.0025$ AU. The stable region visible as black region
diminishes primarily with the distance to the exchange orbit. In the
respective Fig. \ref{ear-i} one can see how these regions start to be limited
with larger inclinations by the 3:1 MMR at $a=0.48$ AU  and the 2:1 MMR at a=0.65 AU. There
exist a decreasingly smaller stable region in between them up to $i = 60^{\circ}$, where the whole region starts to be unstable.
\begin{figure}
\centering
\includegraphics[width=5.5cm,angle=270]{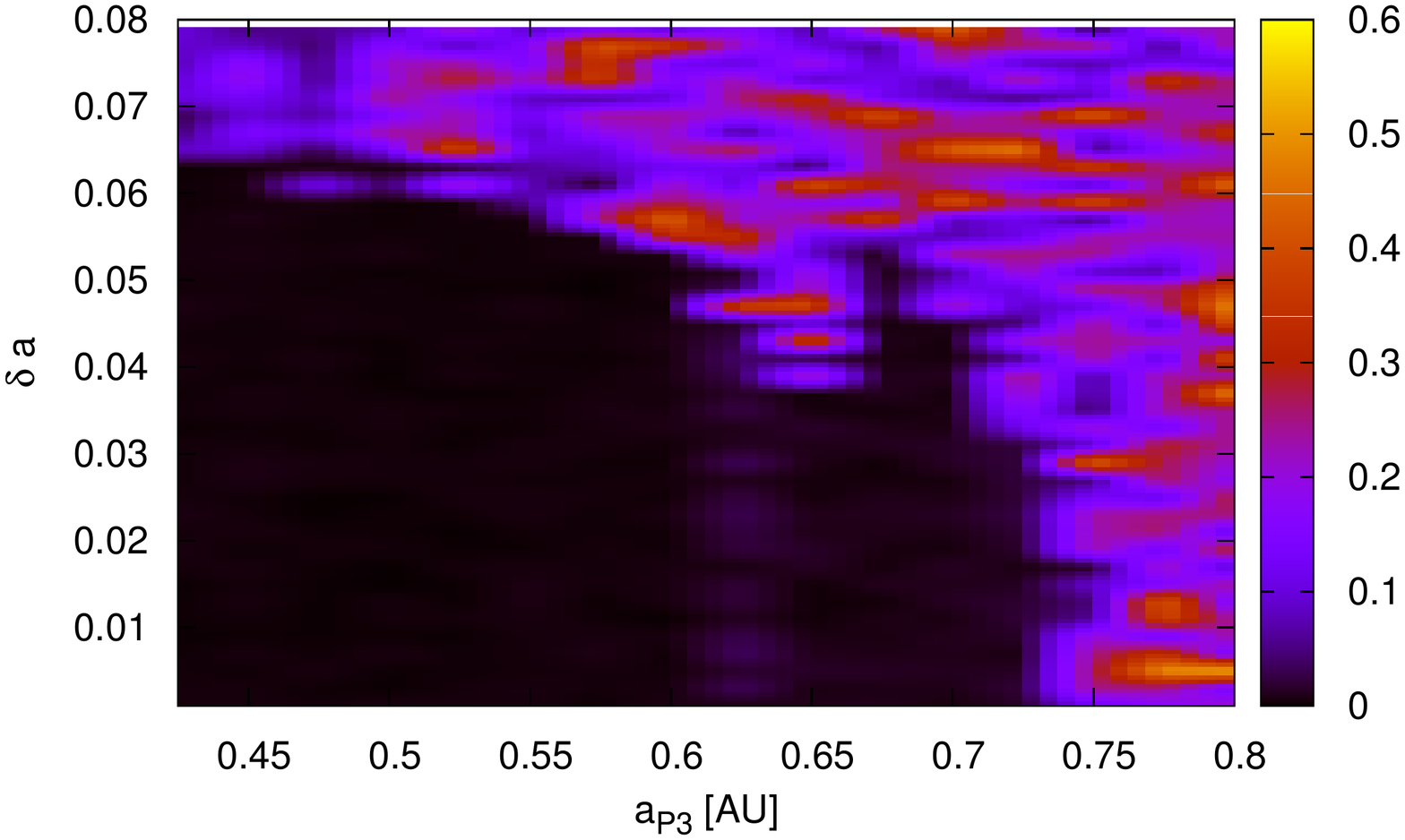}
\includegraphics[width=5.5cm,angle=270]{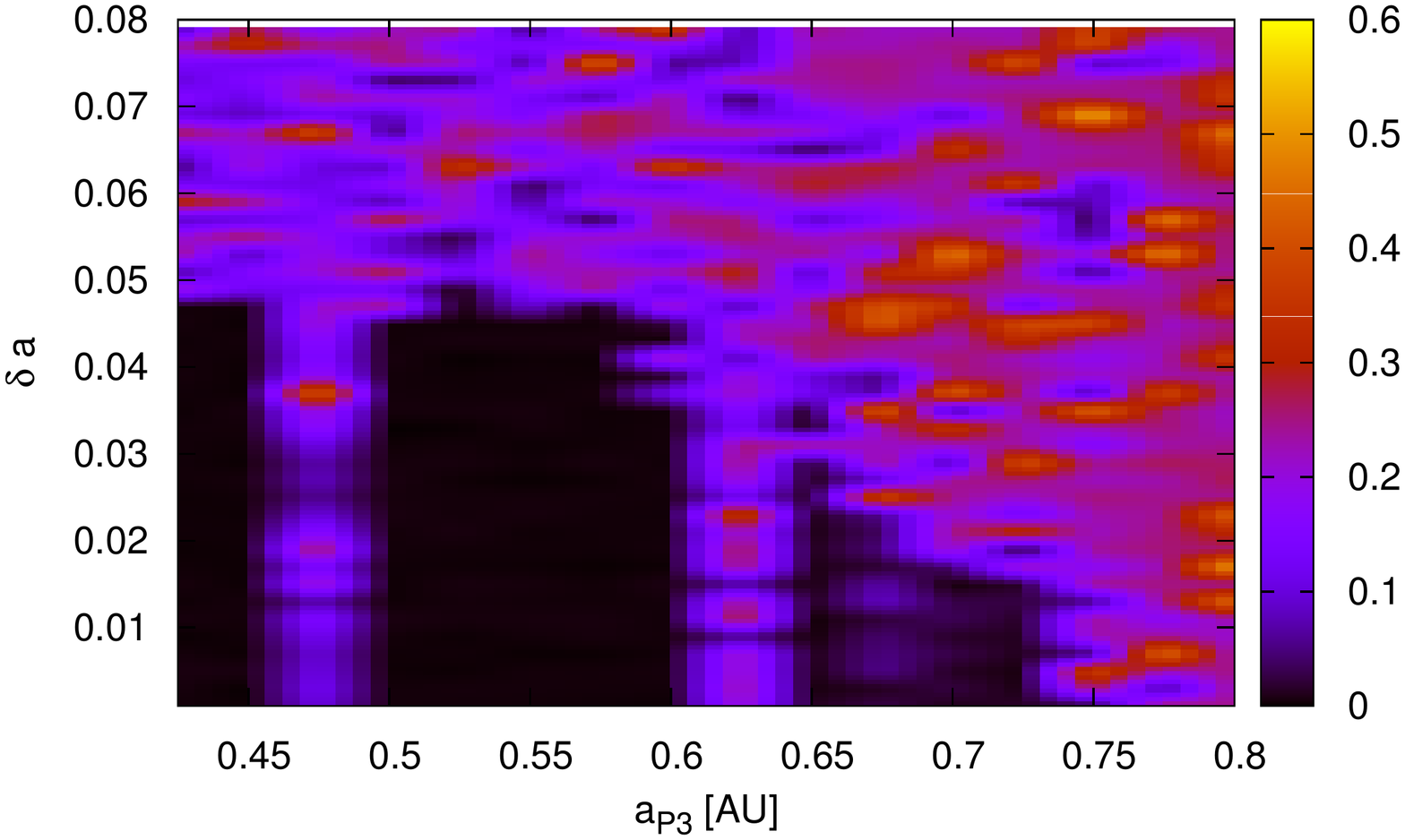}
\includegraphics[width=5.5cm,angle=270]{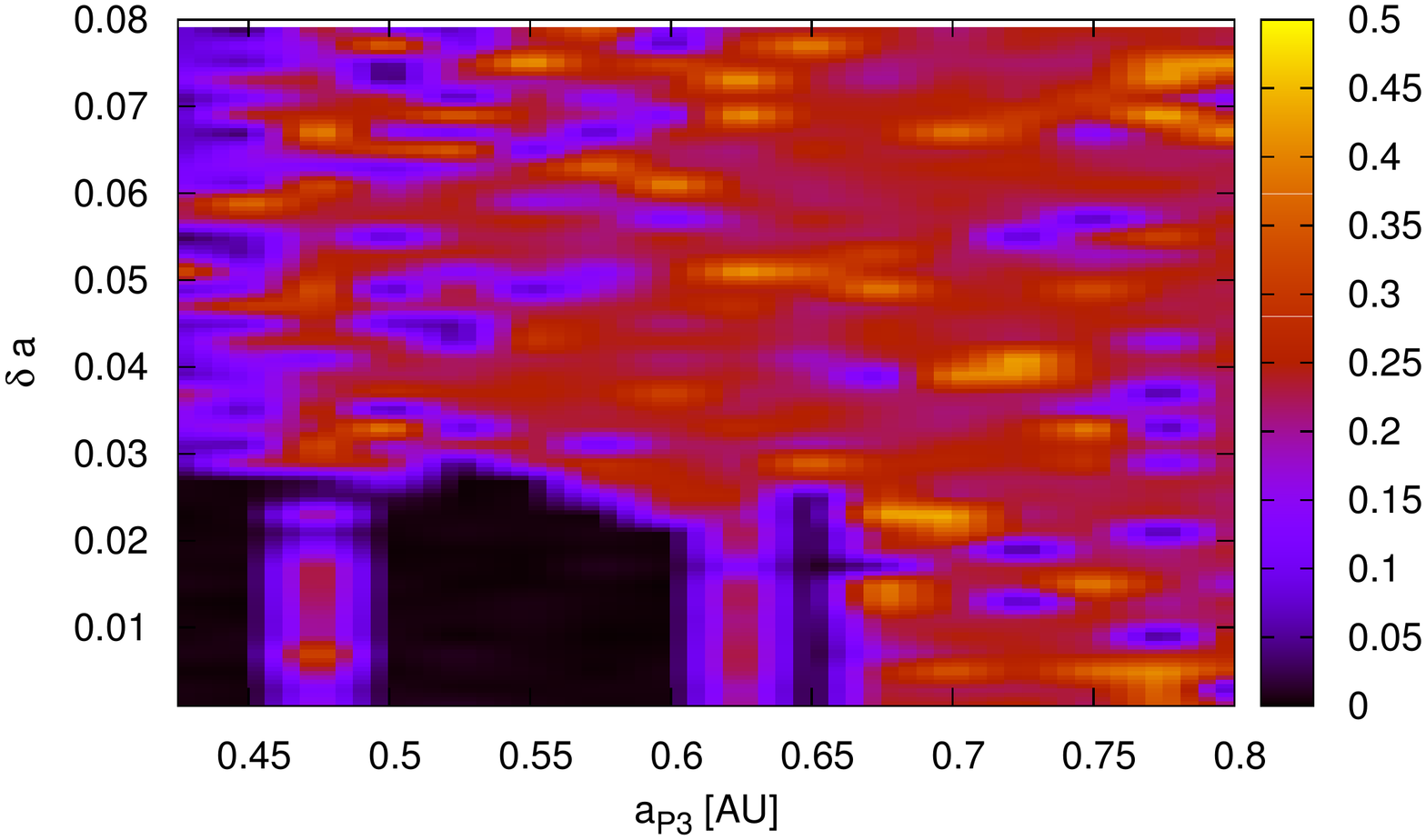}
\caption{The stable region (black) for xch-a orbits under the perturbation
of an inner Jupiter: x-axis position of the planet P3, y-axis initial
difference $\delta a$ in semi-major axes of P1 and P2, z-axis final eccentricity of P1.
The inclination of P2 with respect to the orbits of P1 and P3 is $i=0^{\circ}$ (top graph), $i=16^{\circ}$ (middle graph), and $i=40^{\circ}$ (bottom graph). color code like in
  (Fig.\ref{inclination}}
\label{ear-i}
\end{figure}


\item When P1 and P2 start in the same orbital plane and P3 has an
original inclination with respect to the exchange orbit the picture is not
essentially different, but the stable region in between the two resonances
mentioned above diminishes faster (not shown here).


\subsubsection{Results (c) and (d): an outer Jupiter P3 perturbs P1-P2}

\item In this part of the study of the four-body problem with a perturbing outer
Jupiter, we varied its distance in steps of $\delta a_{P3}=0.085$ for $1.4
\mbox{ AU } < a_{P3} < 4.6$ AU  for eight different inclinations of P1 with respect to the
orbit of P2-P3 for $0^{\circ} < i < 56^{\circ}$. The other initial conditions
were the same as in (a): the stable region is visible as black region
and it diminishes primarily with the decreasing distance to the exchange
orbit. It is obvious (Fig.\ref{ear-o}, left panel) that the stability border is almost
equal with respect to separation in $\delta a$  and diminishes steadily 
with increasing inclination. The inner stability limit is given by the 2:5
MMR (a $\approx$ 1.8 AU) and grows also with larger inclinations. For large
inclinations (from $i= 48^{\circ}$) no xch-a orbits at all can survive.

\begin{figure}
\centering
\includegraphics[width=4.2cm,angle=270]{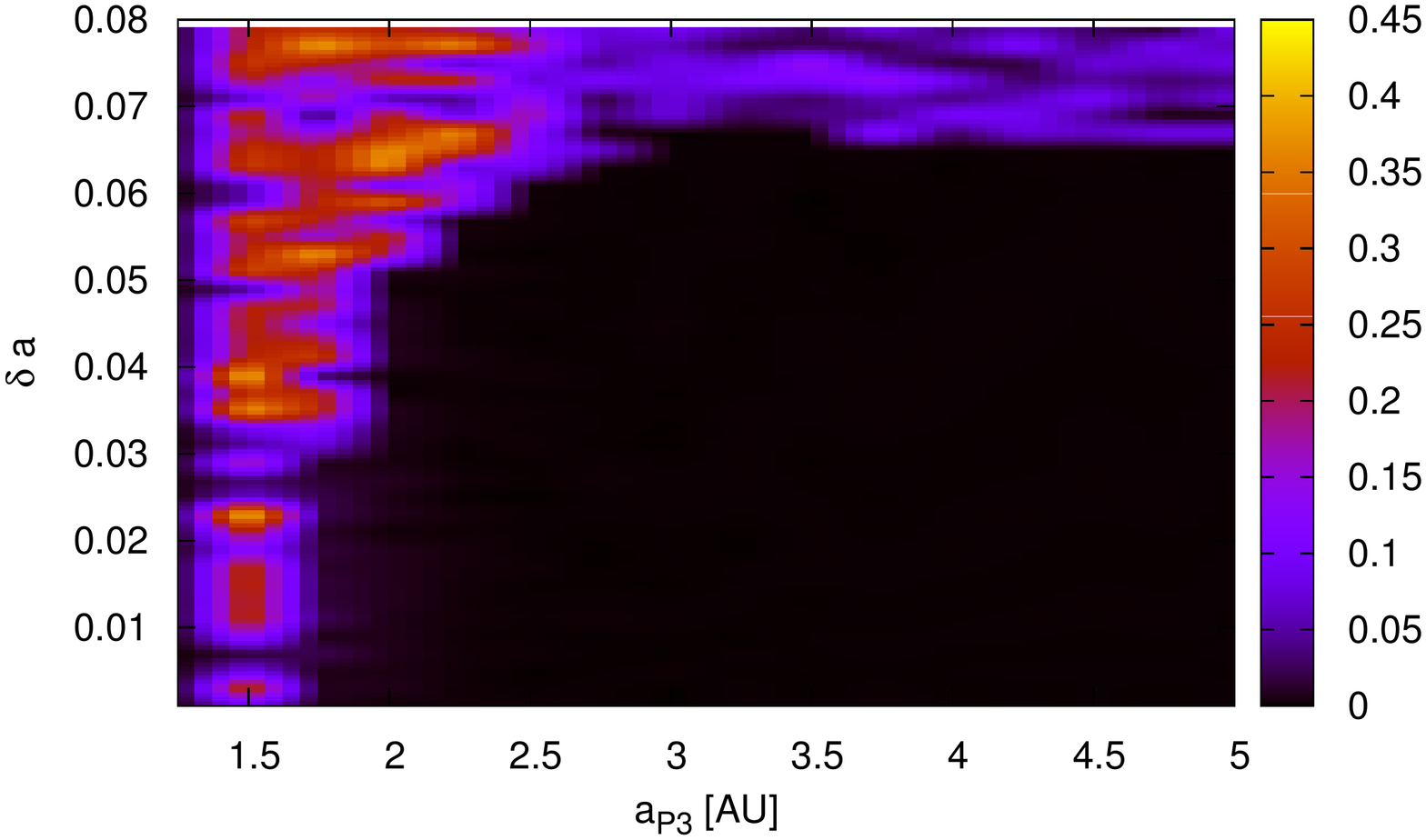}
\includegraphics[width=4.2cm,angle=270]{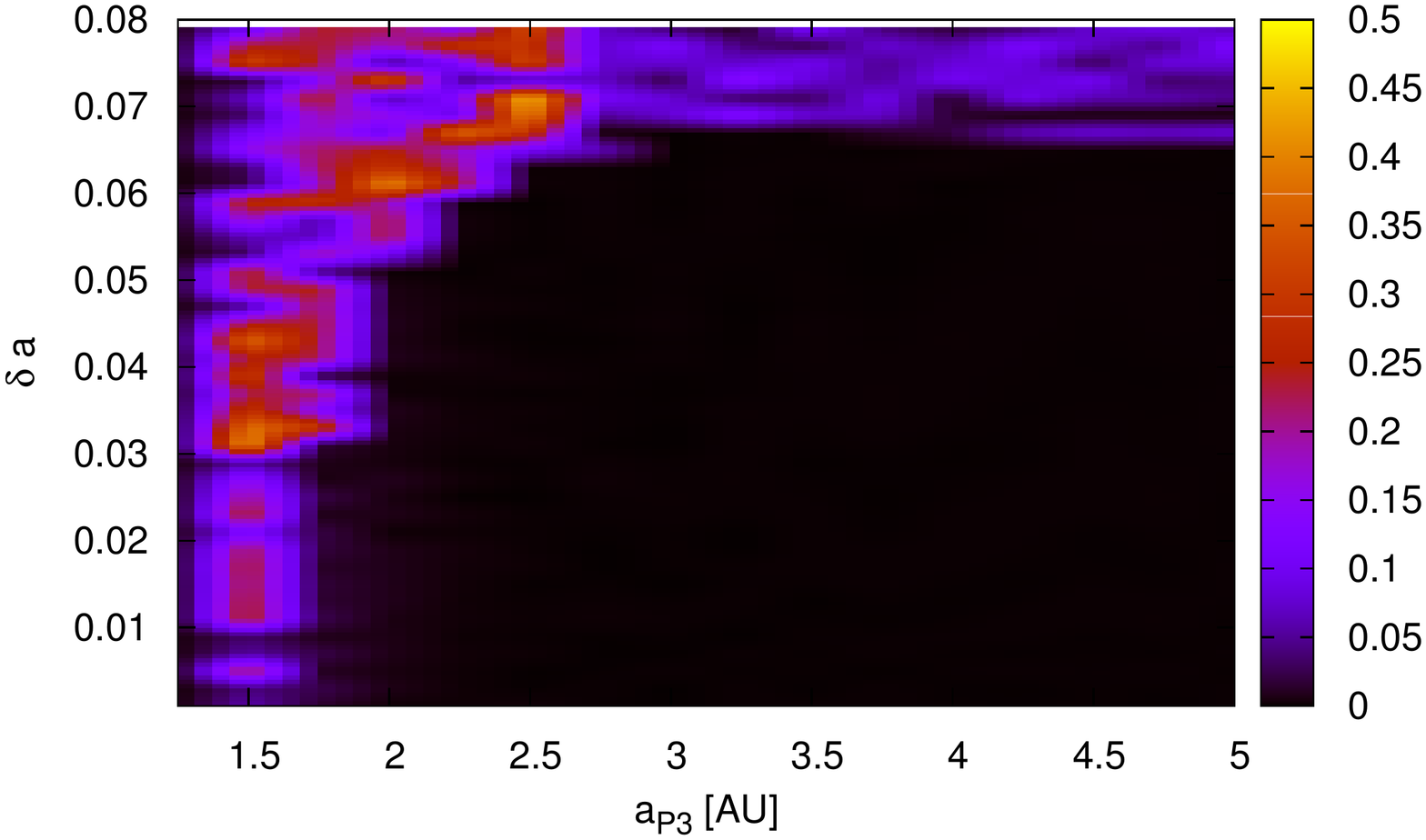}
\includegraphics[width=4.2cm,angle=270]{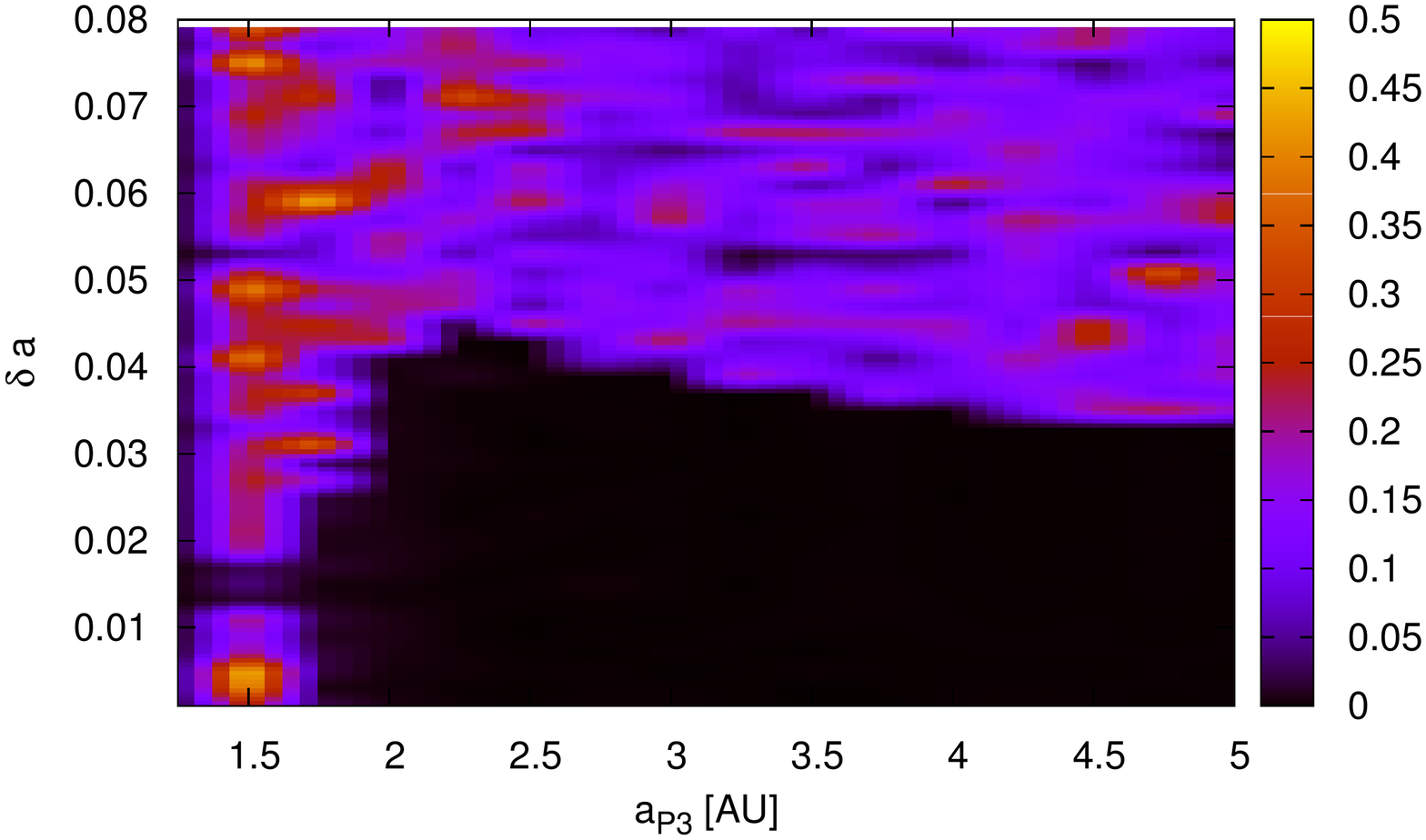}
\includegraphics[width=4.2cm,angle=270]{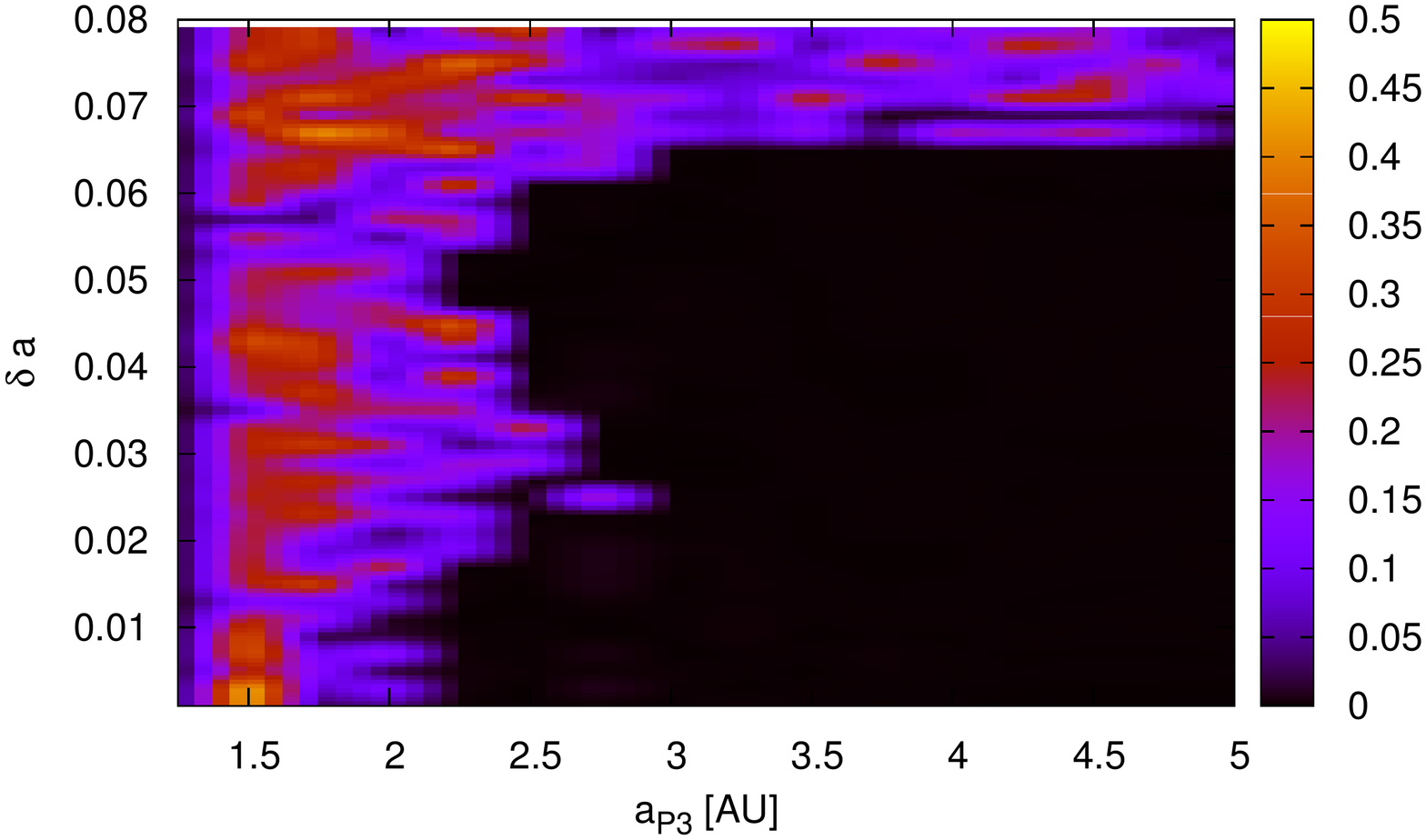}
\includegraphics[width=4.2cm,angle=270]{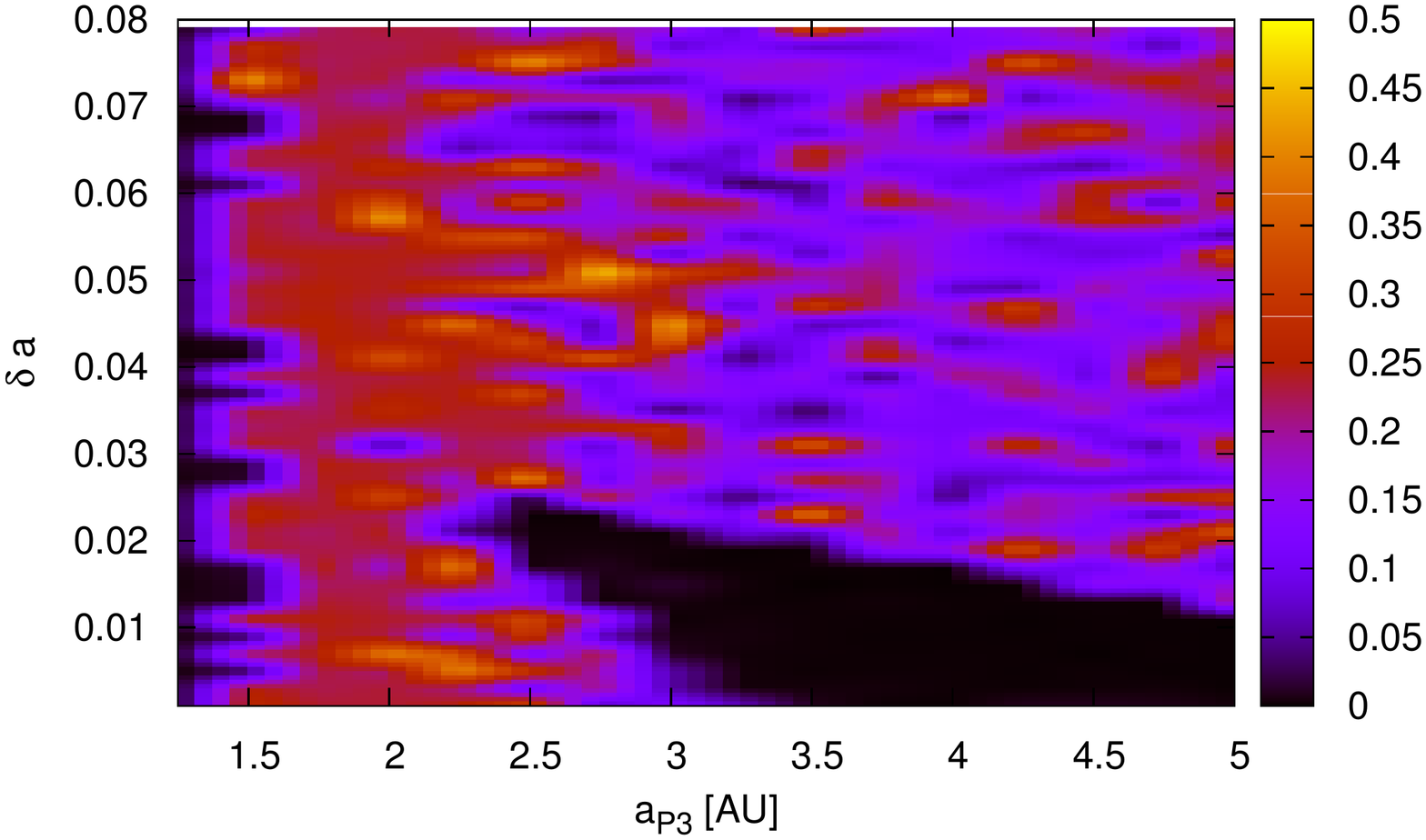}
\includegraphics[width=4.2cm,angle=270]{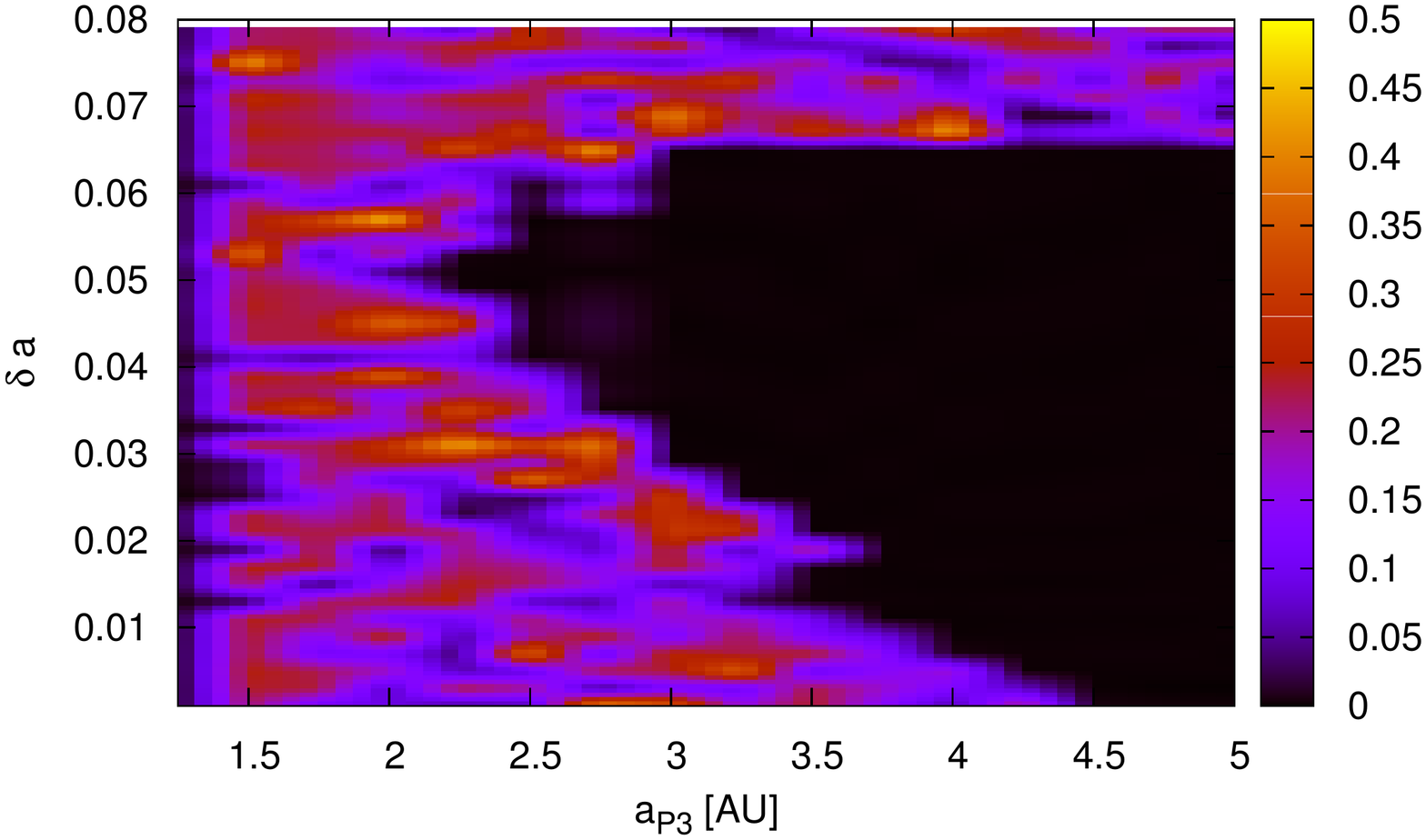}
\caption{The stable region (black) for xch-a orbits under the perturbation
of an outer Jupiter; axes and description like in Fig. \ref{ear-i}. Left panel: the
inclination of P1 is $i=0^{\circ}$ (top graph), $i=16^{\circ}$ (middle graph) and  $i=40^{\circ}$ (bottom graph). Right panel: the inclination of P3 is $i=0^{\circ}$ (top graph), $i=24^{\circ}$ (middle graph) and  $i=32^{\circ}$ (bottom graph).}
\label{ear-o}
\end{figure}

\item We put initially P1 and P2 in the same orbital plane and P3 has an
initial inclination with respect to the exchange orbit. 
As for (c) we varied the distance of P3 with the same step size, but only for six different inclinations for P1 for $0^{\circ} < i < 40^{\circ}$. The stable (black) region 
diminishes as visible in Fig. \ref{ear-o} (right panel) again with the decreasing distance 
to the exchange orbit like in Fig. \ref{ear-o} (left panel) up to  $i = 16^{\circ}$. Then an
interesting -- still unexplained -- feature appears from $i = 24^{\circ}$ on
such that that the stability regions forms a triangle with no stable orbits
for $\delta a$ small. Finally for $i >  40^{\circ}$ in the range of the
perturbations of P3 ($a_{p3}<$ 5 AU) no stable orbits for P1 and P2 survive in the
exchange mode.


\end{enumerate}


\subsubsection{The 3:1 MMR}

In a zoom of the initial condition diagram (Fig.\ref{ear-i}) 
for motion of the perturber P3 in 3:1 MMR with P1 and P2 we
observe the growing of the perturbation with the inclination between the
orbits of P1 and P2 with respect to
$\Delta a$ (Fig. 11). For very small inclinations the resonance is almost invisible but
already with an inclination of $3^{\circ}$ one can see the double structure of
the resonance which may be explained by the difference in distance of P1
(P2) which jump mainly between two different distinct semi-major axes. With larger inclinations 
between P1 and P2 the two unstable areas enlarge -- still
separated -- but from $i=9^{\circ}$ the two unstable regions join. At the same time
the extension of the separation $\Delta a_{P3}$ decreases with increasing inclination. From $i=15^{\circ}$ on there are only small stable areas on both
sides of this MMR. 

\begin{figure}
\begin{center}
\includegraphics[width=5.5cm,angle=270]{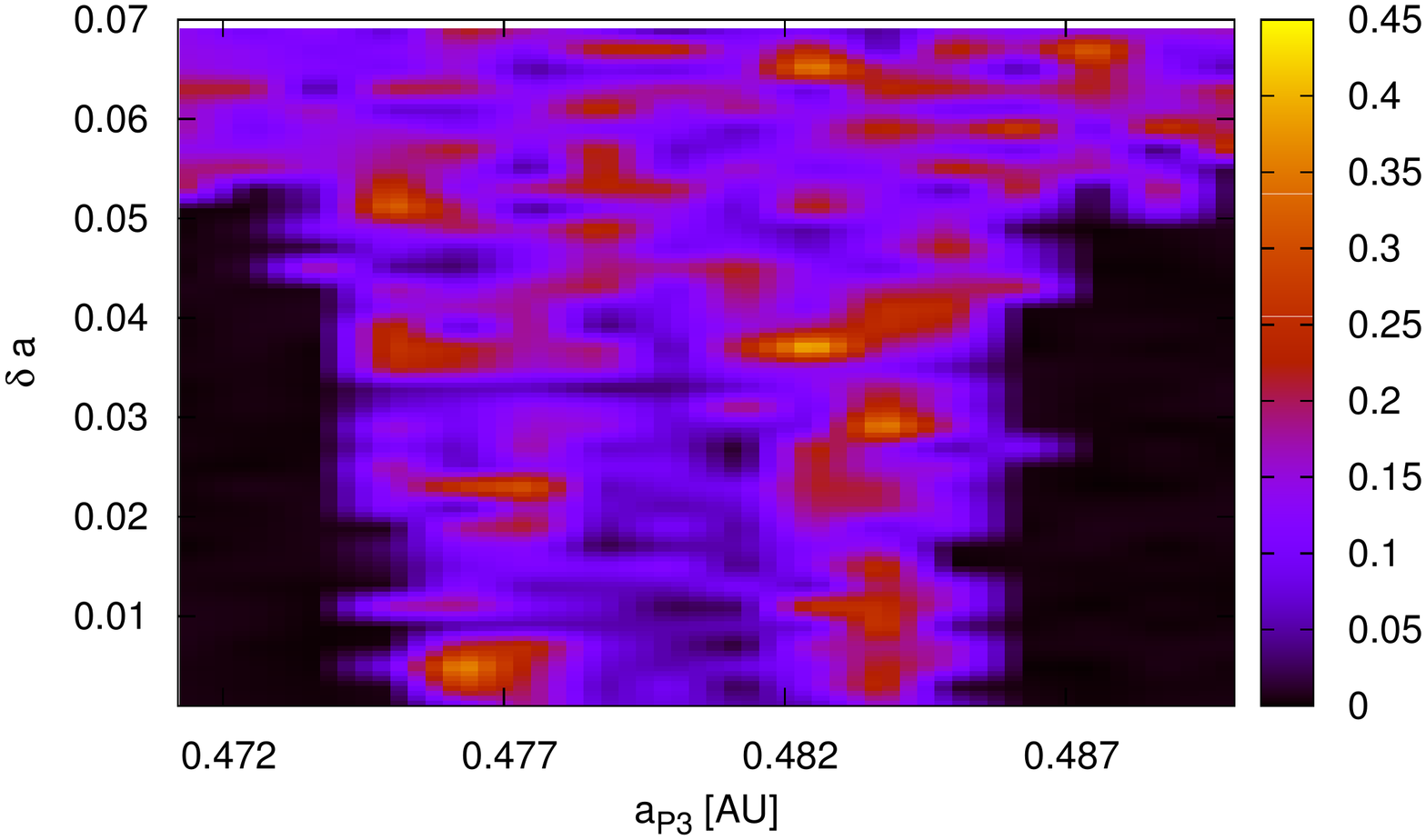}
\includegraphics[width=5.5cm,angle=270]{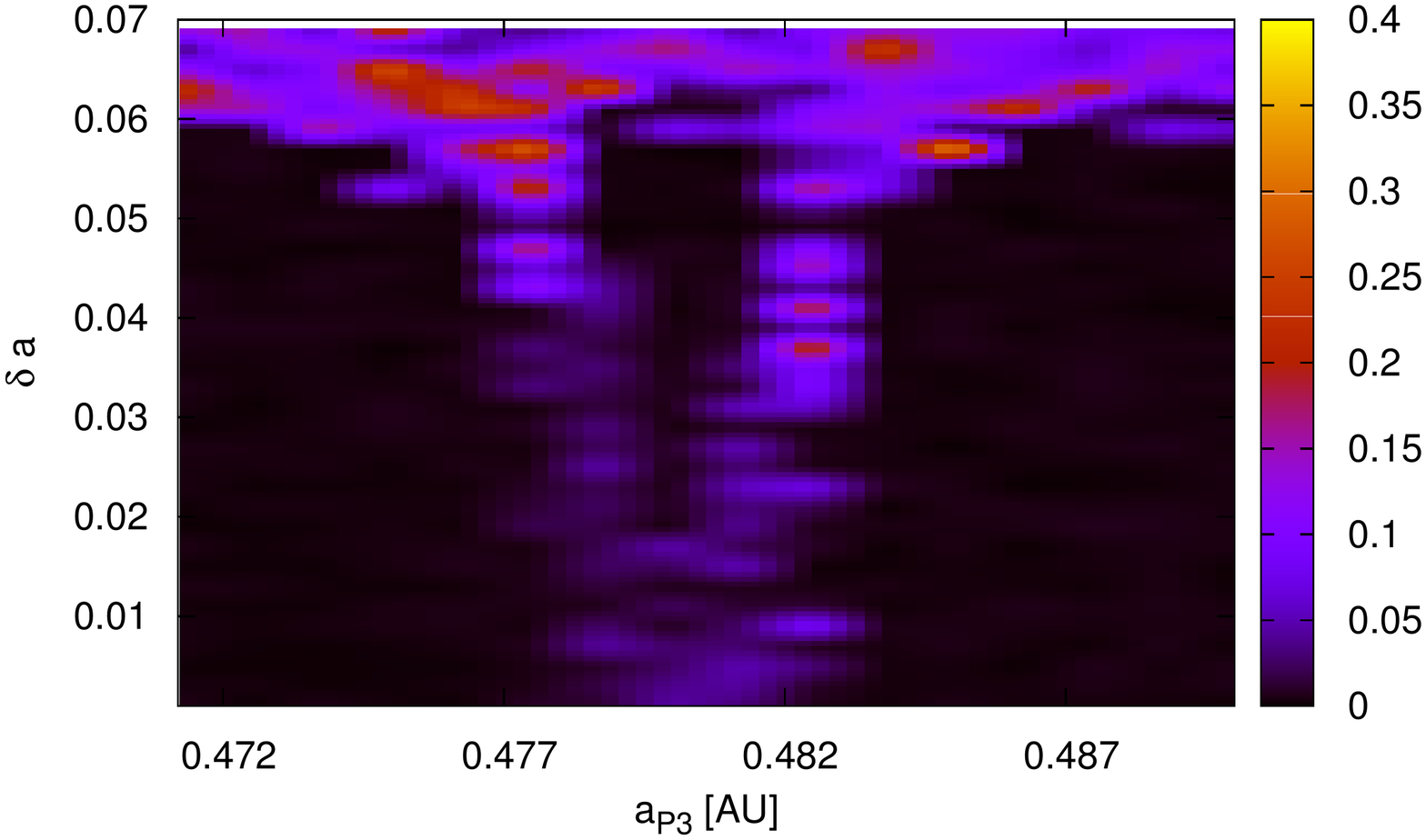}
\caption{Initial condition diagram of the 3:1 MMR Hot Jupiter - two
  Earth-like planets in
  xch-a orbits in 1 AU from a Solar type star: x-axis distance of P3, y-axes initial distance between the two Earth-like planets, z-axis,
maximum eccentricity of the planets in xch-a orbits. The black region
indicate stable exchange orbits. Top graph: i = $3^{\circ}$, bottom graph: i = $9^{\circ}$.}
\end{center}
\label{fig27}
\end{figure}

\subsubsection{The 2:1 MMR}
 In the zoom of the former graphs (Fig. \ref{ear-i}) we can see in
 Fig. 14 the Y-type structure of the
 2:1 MMR for  $i=9^{\circ}$ with increasing initial distance of the P1 and P2
 from each other. This is also the picture we know from resonances that
 according to the semi-major axes the perturbations grow with larger
 eccentricities; here the difference in initial semi-major axes of P1 and P2 is
 playing the role of eccentricity.  

\begin{figure}
\begin{center}
\includegraphics[width=8.0cm,angle=270]{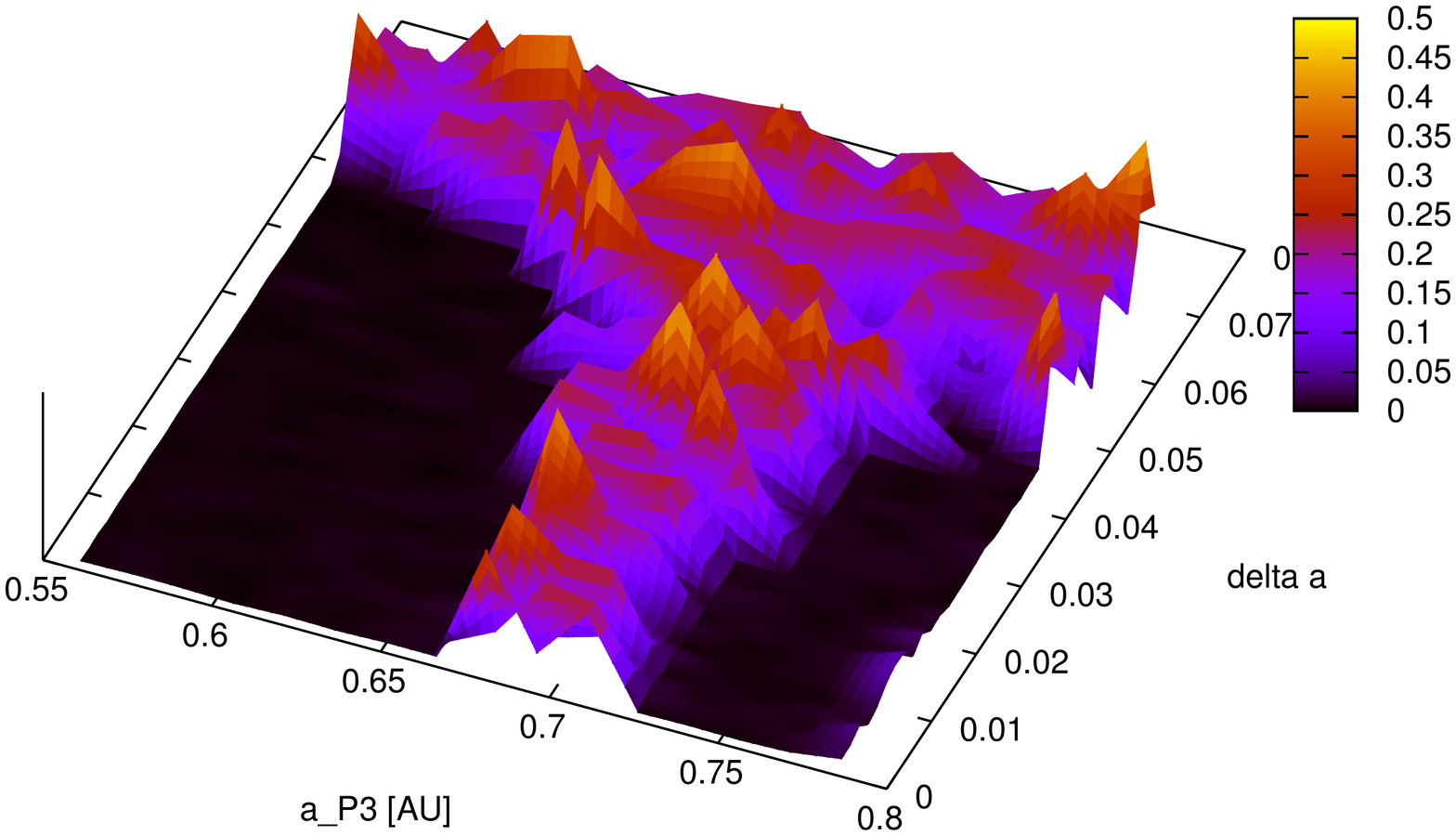}
\caption{Initial condition diagram of the 2:1 MMR; captions like in Fig. \ref{ear-i} but
  only for $i=9^{\circ}$}
\end{center}
\label{fig52}
\end{figure}


\subsection{Summarized results for the xch-a orbits}

\begin{figure}
\begin{center}
\includegraphics[width=1.5in,angle=270]{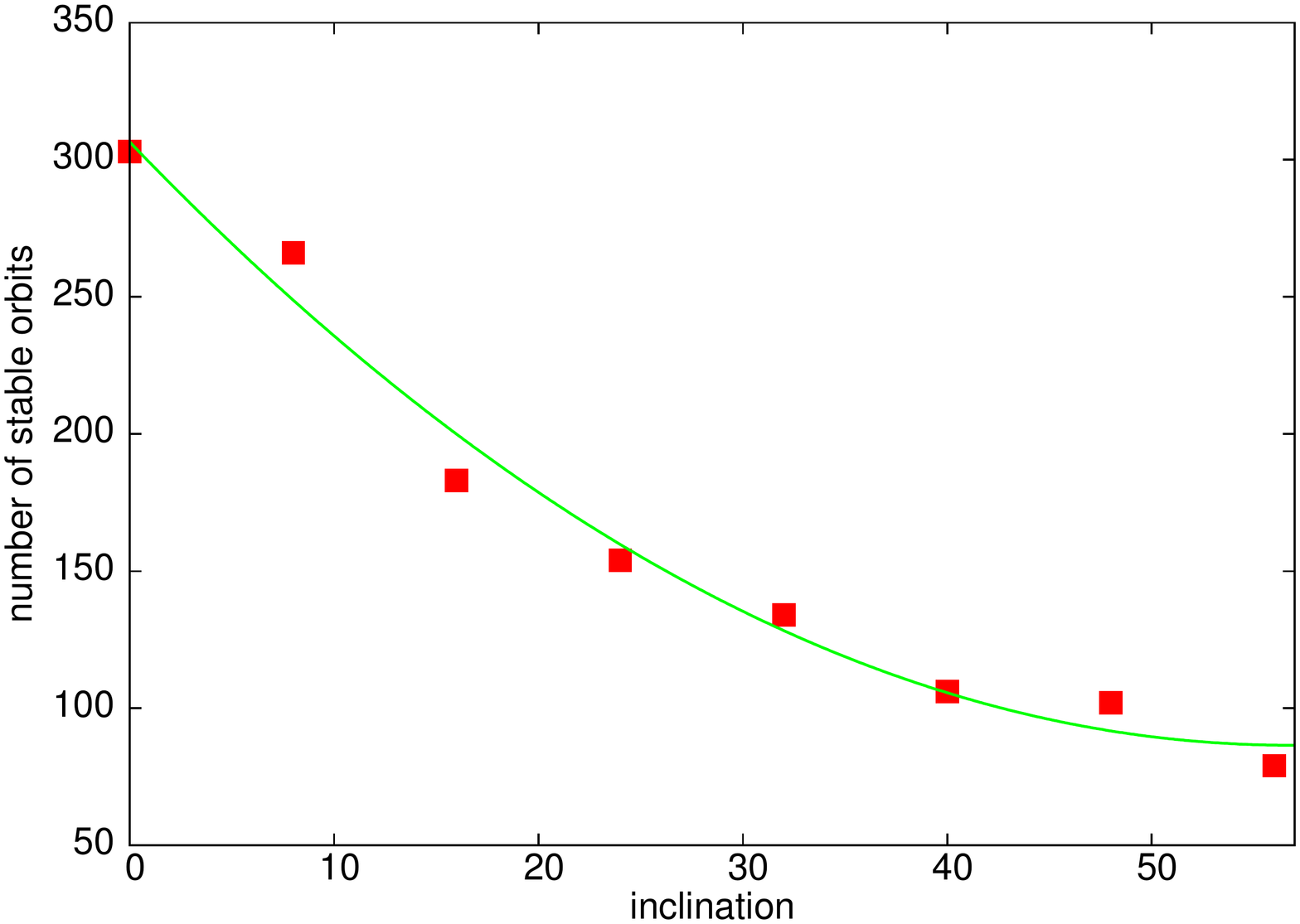}
\includegraphics[width=1.5in,angle=270]{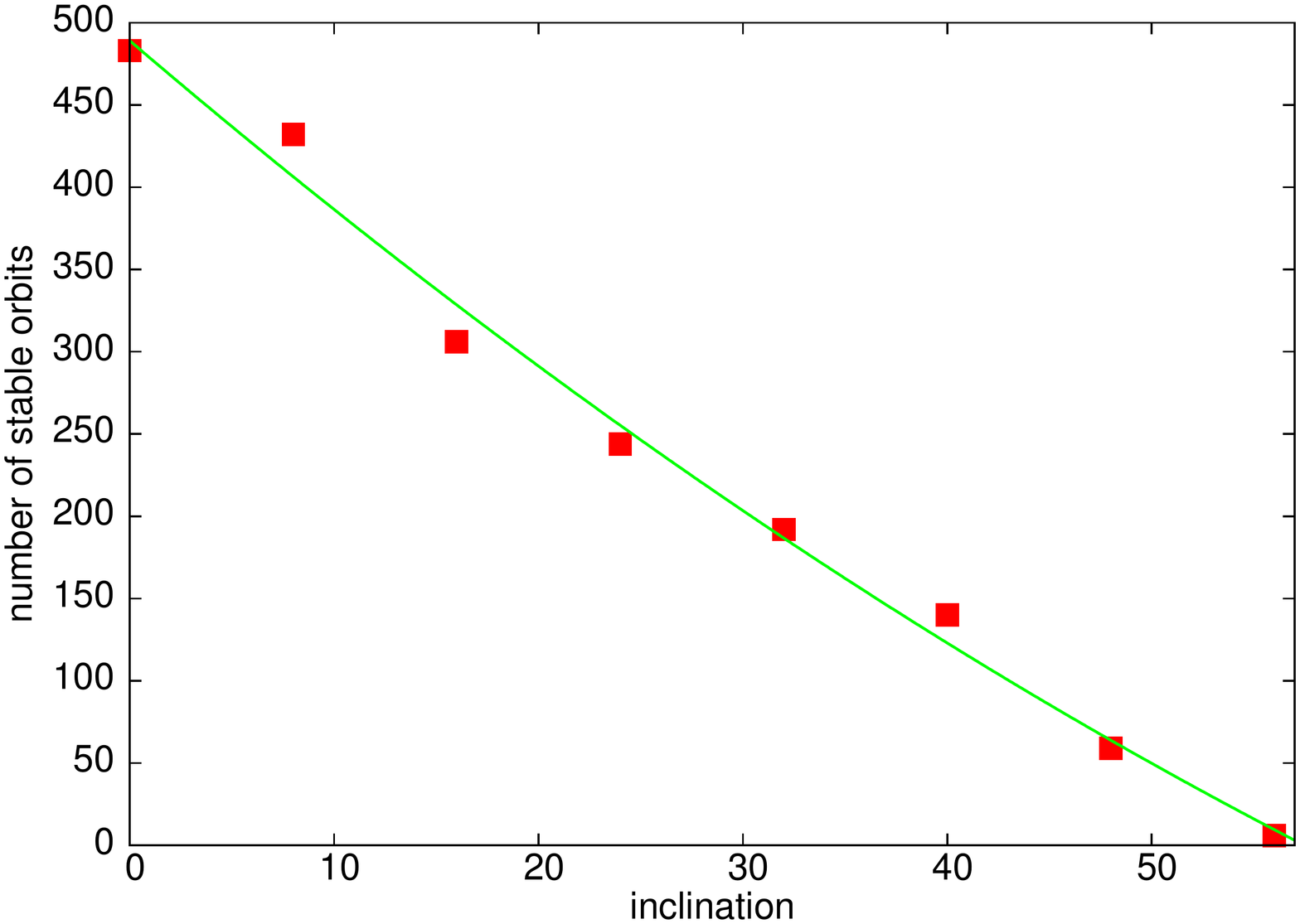}
\includegraphics[width=1.5in,angle=270]{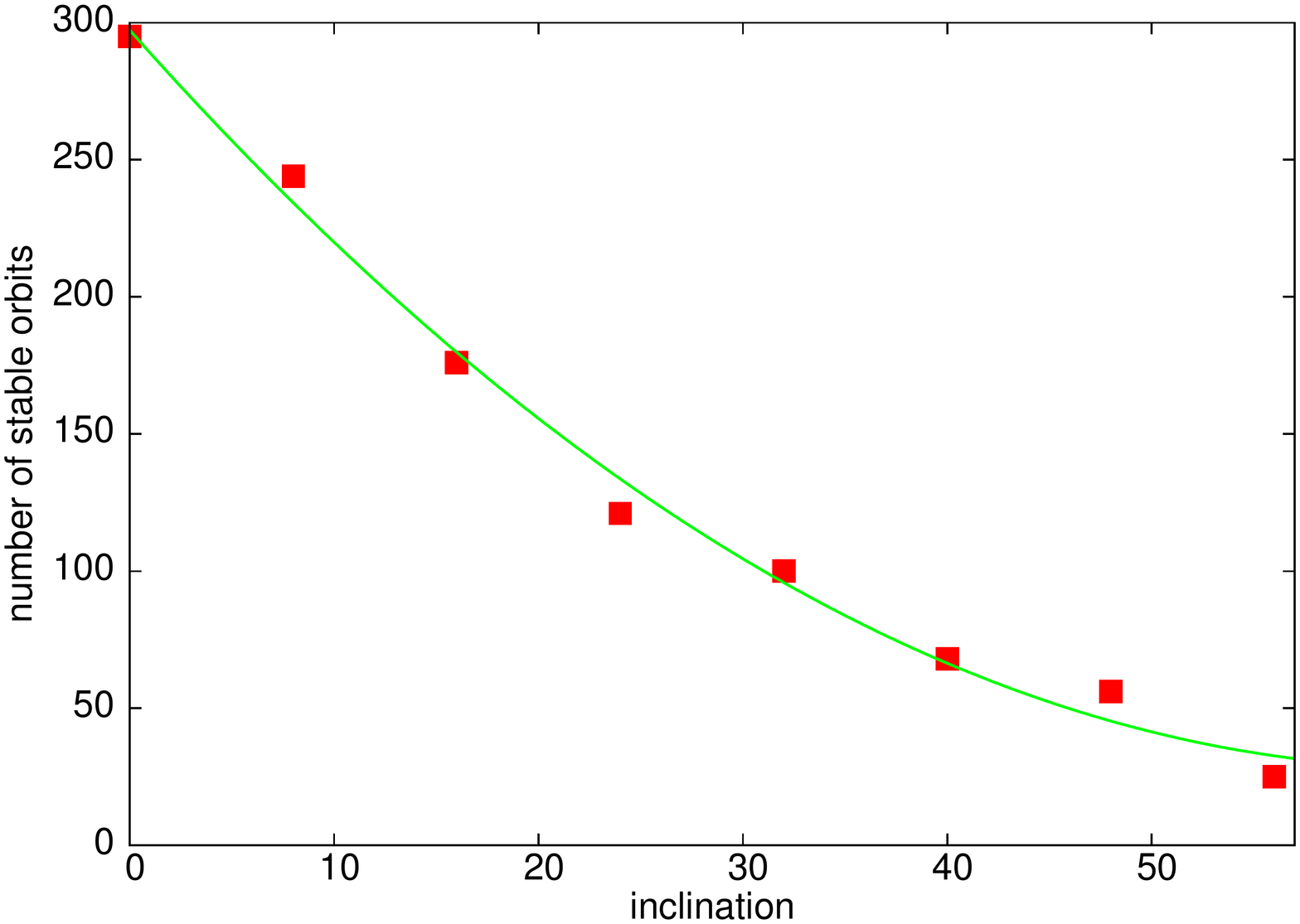}
\includegraphics[width=1.5in,angle=270]{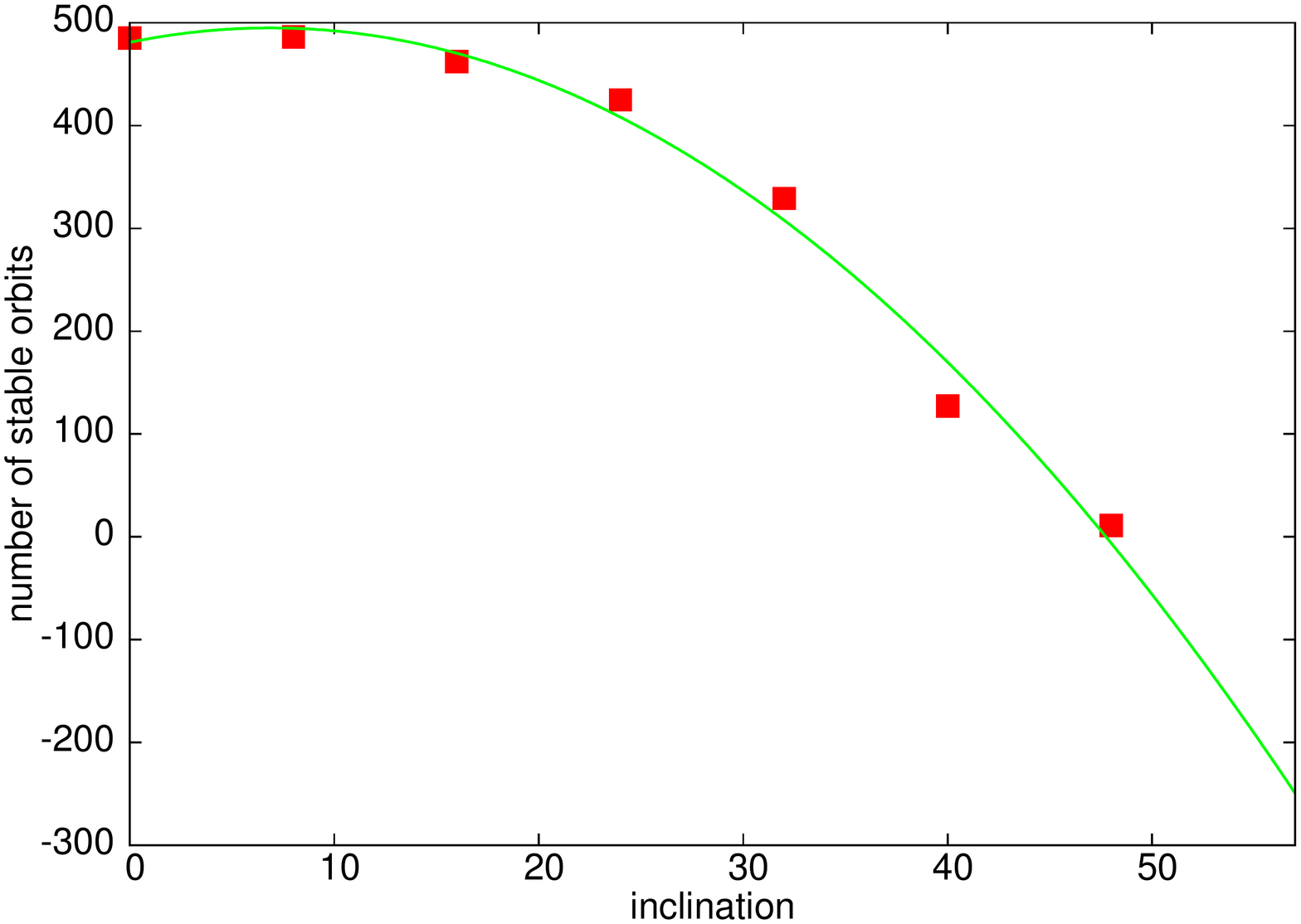}
\caption{Best fits for the data of cases (a) (Fig. \ref{ear-i}), (b), (c) (Fig. \ref{ear-o}, left panel) and (d) (Fig. \ref{ear-o}, right panel); x-axes inclination of one planet with respect to the same orbital plane of the two other ones, y-axis number of stable orbits. left, upper graph: P2 inclined, inner perturber; right, upper graph: P2 inclined, outer perturber; left, bottom graph: P3 inclined, inner perturber; right , bottom graph: P3 inclined, outer perturber.}
\end{center}
\label{ear-jup-io}
\end{figure}

It is well visible that in all four cases (a) - (d) (Fig. 14) the
dependence of the largeness (number of stable orbits) with inclination is well
represented by a parabola ($f(i) = a \cdot i^2 + b \cdot i + c$), where the values for a, b and c are given in table \ref{coef}.
\begin{table}
\begin{center}
\caption{Parameters of the values a, b and c in the parabola fits given in Fig. 14.}
\begin{tabular}[h]{|l|ccc|}
\hline
case &  a & b & c\\
\hline
(a)& $0.0683594 \pm 0.01545$ & $-7.75521 \pm 0.8996$ & $306.458 \pm 10.78$\\
(b)& $0.0368304 \pm 0.02208$ & $-10.625 \pm 1.286$ & $489 \pm 15.42$\\
(c)& $0.0653832 \pm 0.01162$ & $-8.38616 \pm 0.6767$ & $297.208 \pm 8.111$\\
(d)& $-0.296317 \pm 0.04684$ & $4.07589 \pm 2.34$ & $480.857 \pm 23.98$\\
\hline
\end{tabular}
\label{coef}
\end{center}
\end{table}
For the xch-a orbits perturbed from an inner 'Jupiter' the
fits are very similar (left panel); for an outer perturber (right panel) the difference is quite large
whether the earth-like planets orbit initially in the same orbital plane (left
graph) or they start with different inclinations. In the second case the
disappearance is quite abrupt.

\section{EXCHANGE-E ORBITS}
 
\begin{figure}
\centering
\includegraphics[width=7.0cm]{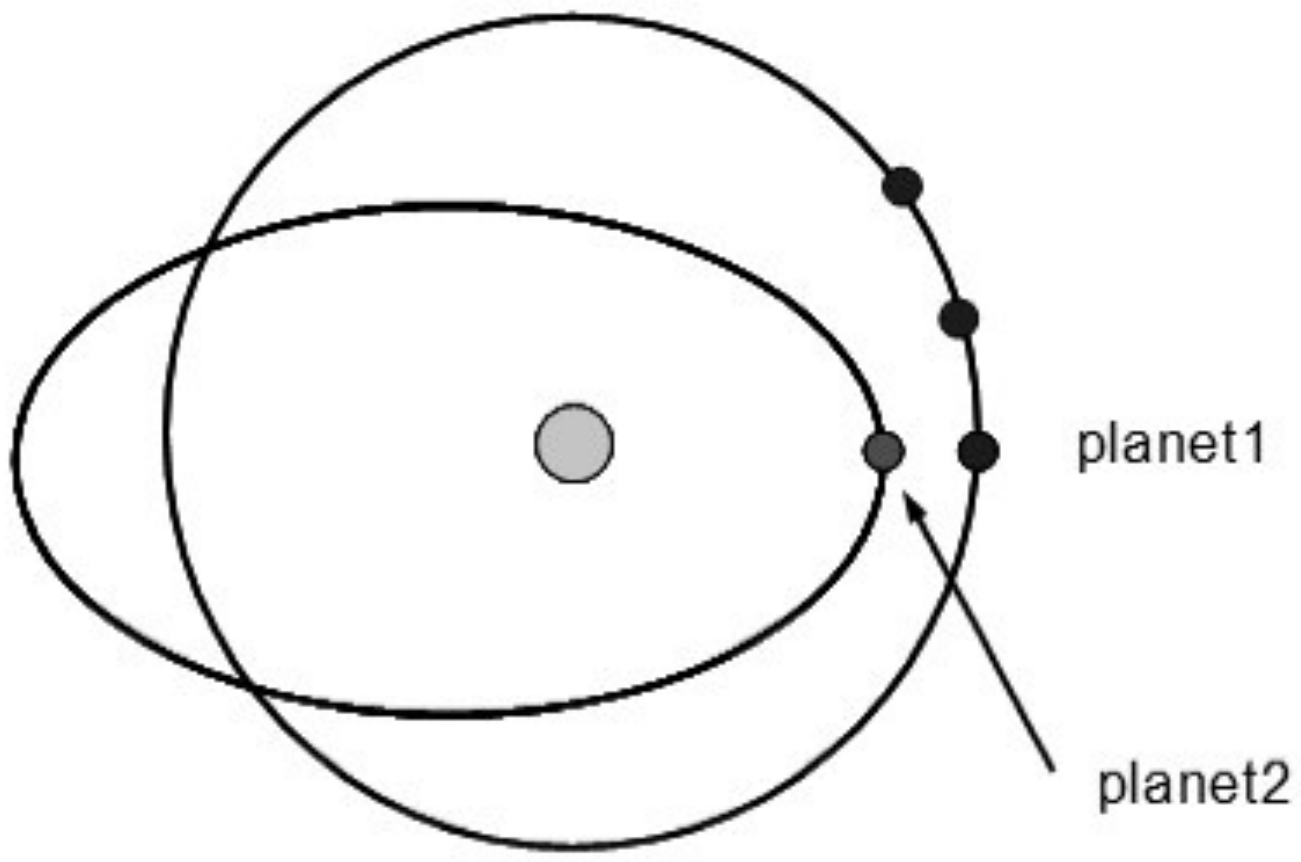}
\caption{Schematic graph for the initial xch-e configuration: P1 is always 
started on a circular orbit with different mean anomalies, while P2 is initially  on 
different eccentric orbits with a zero mean anomaly.}  
\label{initcon}
\end{figure}

The xch-e configuration: Two planets are moving on nearly the same orbit, one on a 
circular orbit and the other on an eccentric one. The first planet is displaced by the mean anomaly (M), 
which is shown in Fig. \ref{initcon}. Computations have shown that this configuration is also
  longtime stable like the Trojan configuration (Funk et al., 2011). We call it xch-e configuration, because during the
  motion of the massive planets they 'exchange' their eccentricity. The period of such an exchange 
depends on the masses of the planets and the initial eccentricity of the first planet.\\
Numerical studies concerning the xch-e orbits were done by Roth (2009) and Nauenberg (2002). He discussed the possibilities to detect such orbits in extrasolar planetary systems and showed that with the aid of RV measurements xch-e orbits could be distinguishable from the case of a single planet and that they are long-term stable. Funk et al. (2011) focused on the xch-e configuration and determined the stable regions in dependence on the mean anomaly and the eccentricity of the planets. With the help of least square fits they defined the borders of the 
stable region for each mass ratio ($\mu$)\footnote{$\mu$ = $M_{Planet}/(M_{Planet}+M_{Star})$, where $M_{Planet}$ = $M_{Planet1}$ = $M_{Planet2}$} of the lower mass region and have shown that 
the stable region shrinks with larger $\mu$ and disappears completely at $\mu > 0.019$. Furthermore the results show that the xch-e
configurations remain stable for eccentricities up to $e=0.9$.\\
For the present study we fixed the mass ratio $\mu$ and the mean anomaly (M) and extended the parameter space by including the influence of the inclination $i$. In a further step we added a perturbing body and tested its influence on the planets in xch-e motion. A detailed description of the used models, methods and initial conditions is given in Section \ref{modell-e}

\subsection{The dynamical model and the methods}
\label{modell-e}
We performed integrations in the full 3-body and the full 4-body problem, consisting of a host star with 1 $M_{Sun}$, the two planets (P1, P2) in xch-e motion (with 1 $M_{Earth}$ each), and in the case of a perturber a third planet ($m_3$ = 1 $M_{Jupiter}$, P3) moving inside of the two planets in xch-e orbit.
Again all calculations were done by numerical integrations with the Lie integration method (see Hanslmeier \& Dvorak, 1984, Lichtenegger, 1984, Dvorak \& Freistetter, 2005 and Eggl \& Dvorak, 2010), which is capable to deal with high eccentric orbits and close encounters between the bodies.\\
We divided our study into two main cases:
\begin{enumerate}
\item  the unperturbed case (full 3-body problem) integrated for $10^{6}$ years.
\item the perturbed case (full 4-body problem) with a perturbing body between the star and the two planets in xch-e orbits integrated for $10^{5}$ years.
\end{enumerate}
In Table~\ref{init} we show in detail the initial conditions
for both cases.
\begin{table}
\begin{center}
\caption{Initial conditions for the two planets in the xch-e
configuration and for the perturbing planet. Upper part: unperturbed case; lower part: perturbed case. The mass
  of the star was always 1 $M_{Sun}$.}
\begin{tabular}[h]{|l|lll|}
\hline
&  P1 & P2 & P3\\
\hline
a [AU] & 1.0 & 1.0 & -\\
e & 0.0 & 0.02 - 0.9, $\Delta$e = 0.02 & -\\
i [deg] & 0.0 & 0 - 60 & -\\
$\omega$, $\Omega$ [deg] & 0.0 & 0.0 & -\\
M [deg] & 0 & 0 &-\\
Mass & 1 $M_{Earth}$ & 1 $M_{Earth}$ & -\\
\hline
a [AU] & 1.0 & 1.0 & 0.1 - 1.0, $\Delta$a = 0.02\\
e & 0.0 & 0.02 - 0.7, $\Delta$e = 0.02 & 0.0\\
i [deg] & 0.0 & 0.0 & 0 - 56 \\
$\omega$, $\Omega$ [deg] & 0.0 & 0.0 & 0.0\\
M [deg] & 0 & 0 & 0\\
Mass & 1 $M_{Earth}$ & 1 $M_{Earth}$ & 1 $M_{Jupiter}$\\
\hline
\end{tabular}
\label{init}
\end{center}
\end{table}
As stability parameter we used the deviation $\Delta a$ from the initial semi-major axes. It turned out that $\Delta a < 0.001$ was as good stability criterion (for a Detailed description see Funk et al., 2011). In section \ref{unperturbed-e} we show the results for the unperturbed case and in section \ref{planar} the results in the planar perturbed case.

\subsection{The unperturbed case}
\label{unperturbed-e}
\begin{figure}
\centering
\includegraphics[width=7cm,angle=270]{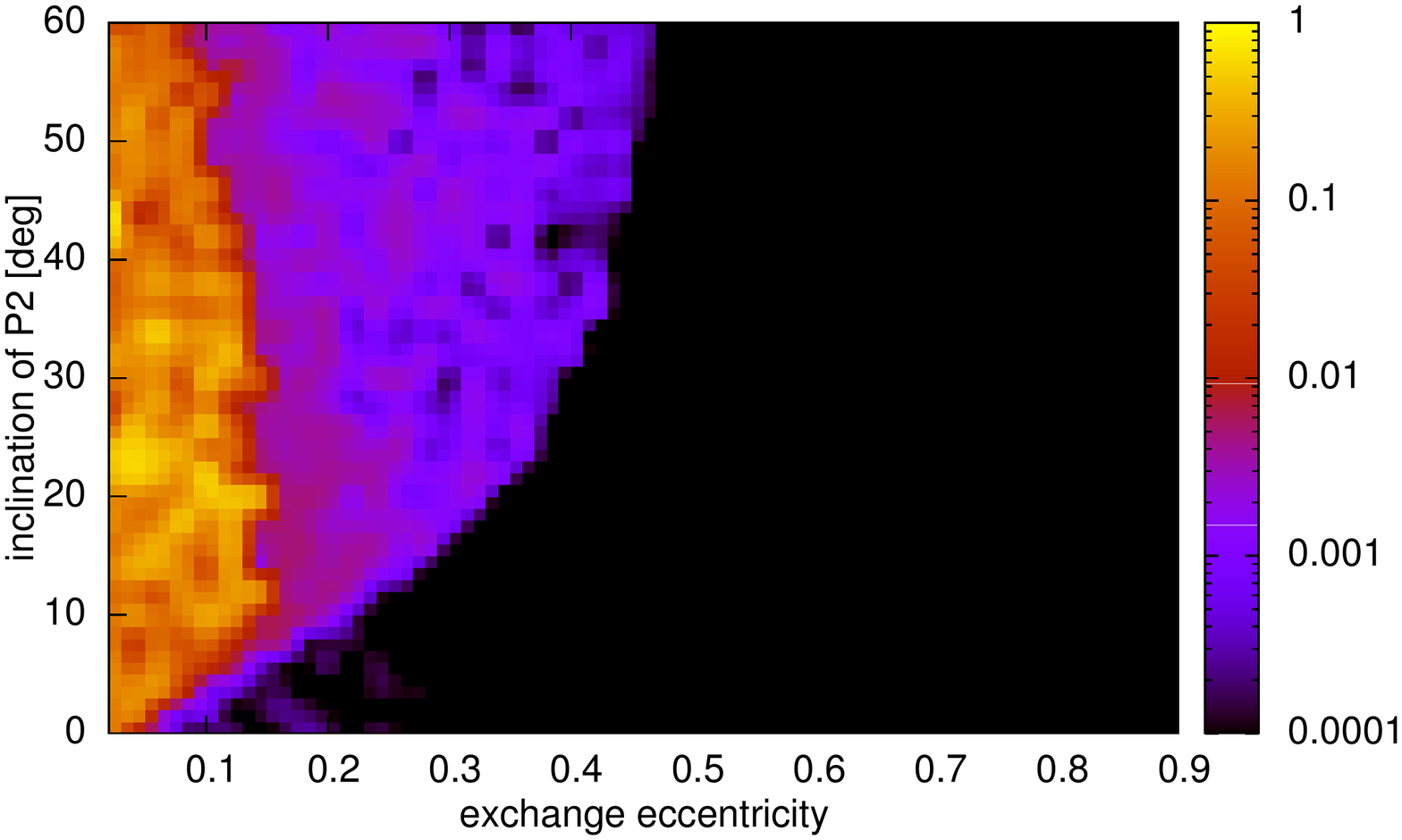}
\caption{Stable regions of xch-e orbits, where on the x-axis we show the exchange eccentricity and on the y-axis the inclination of P2. The color code gives the $\Delta a$ values, where black shows stable motion.}
\label{unper}
\end{figure}

\begin{figure}
\centering
\includegraphics[width=4.2cm, angle=270]{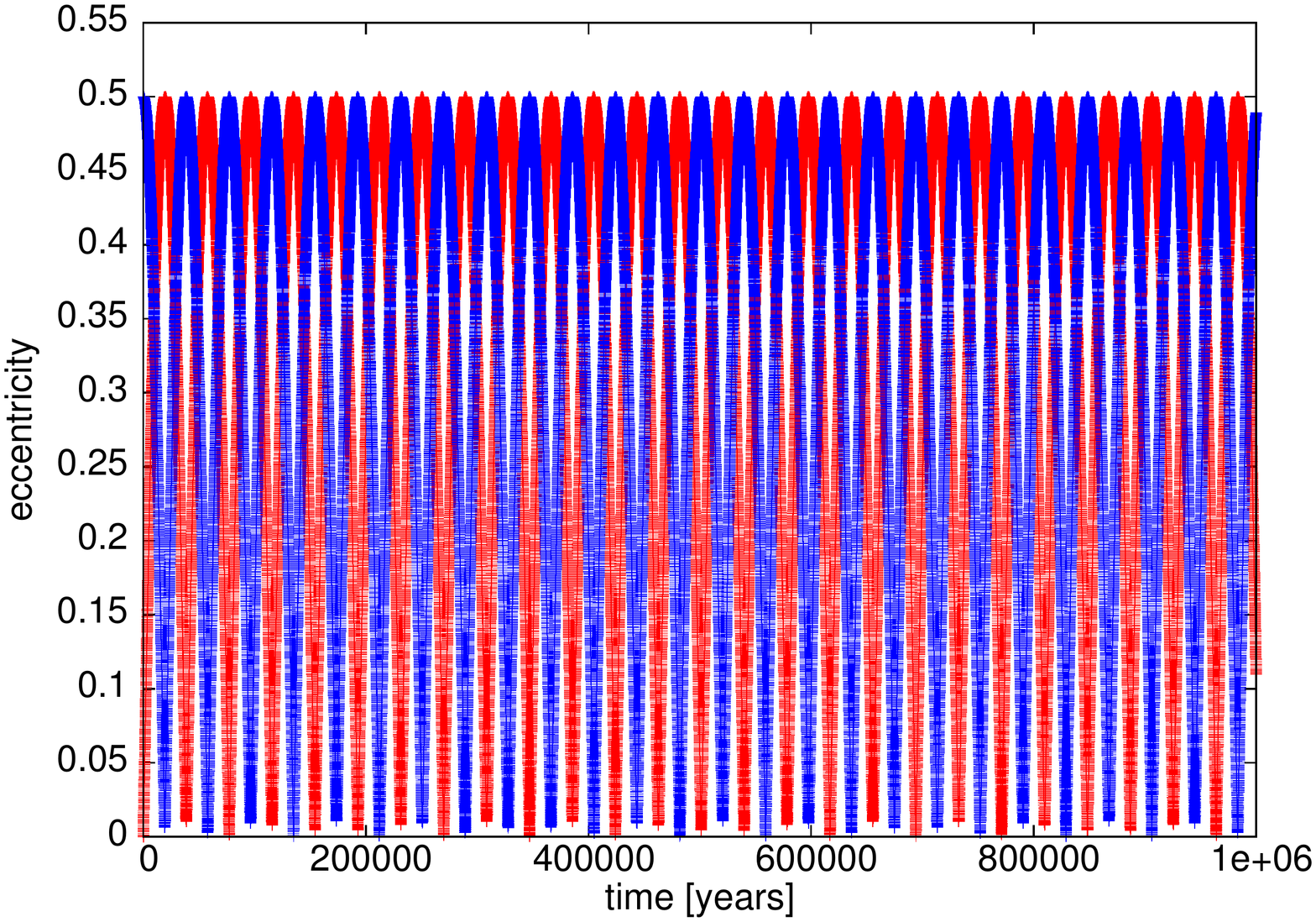}
\includegraphics[width=4.2cm, angle=270]{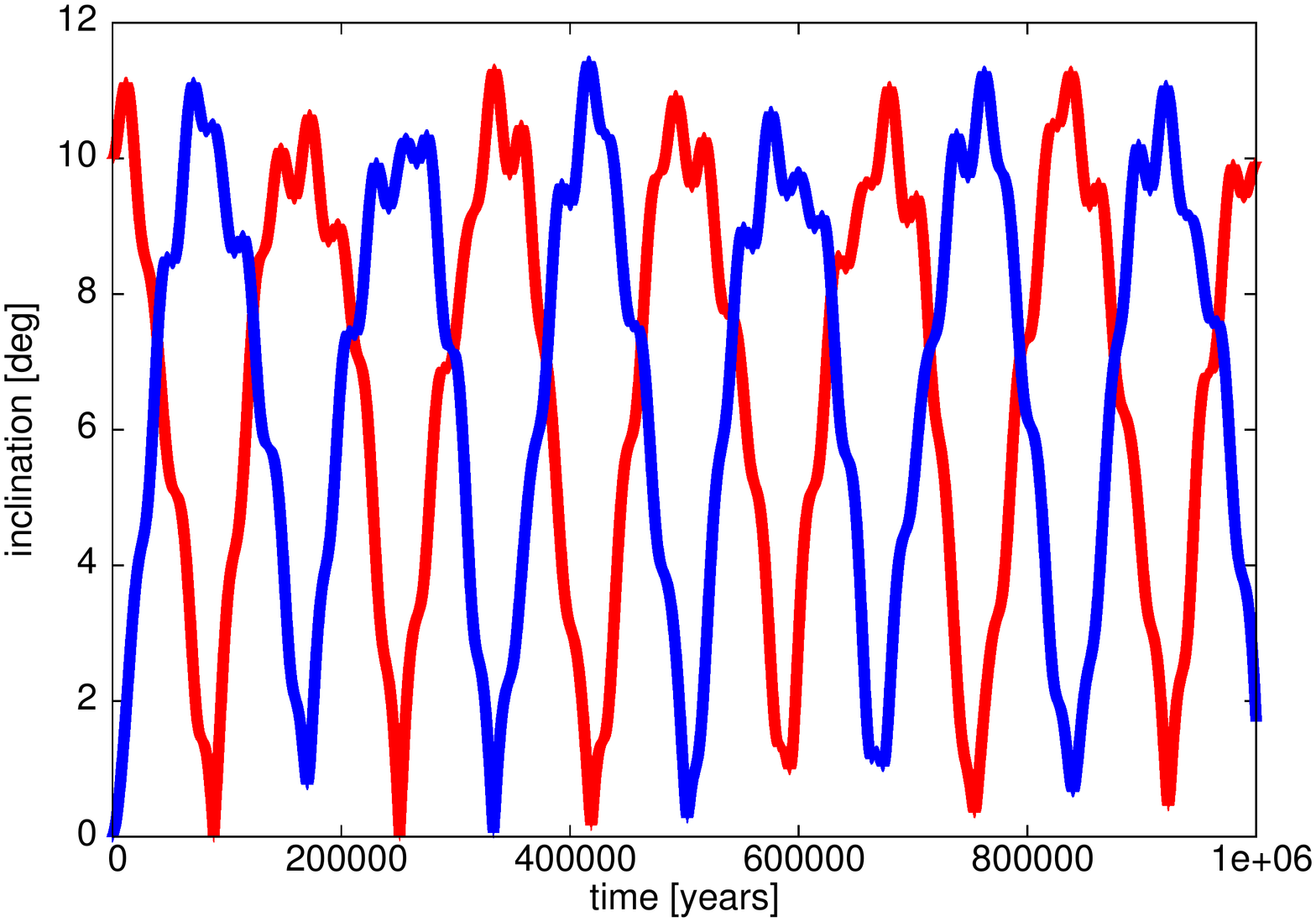}
\includegraphics[width=4.2cm, angle=270]{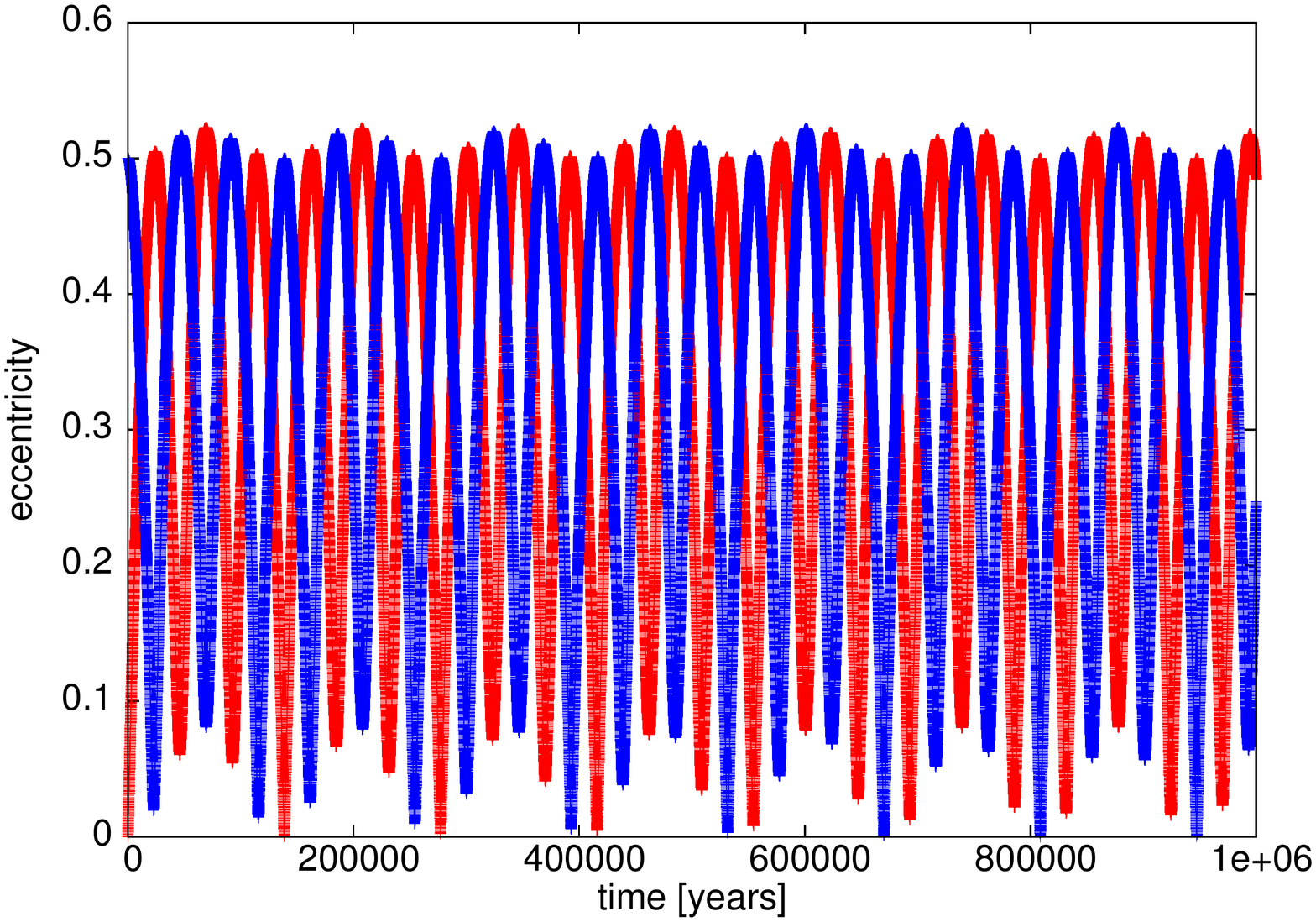}
\includegraphics[width=4.2cm, angle=270]{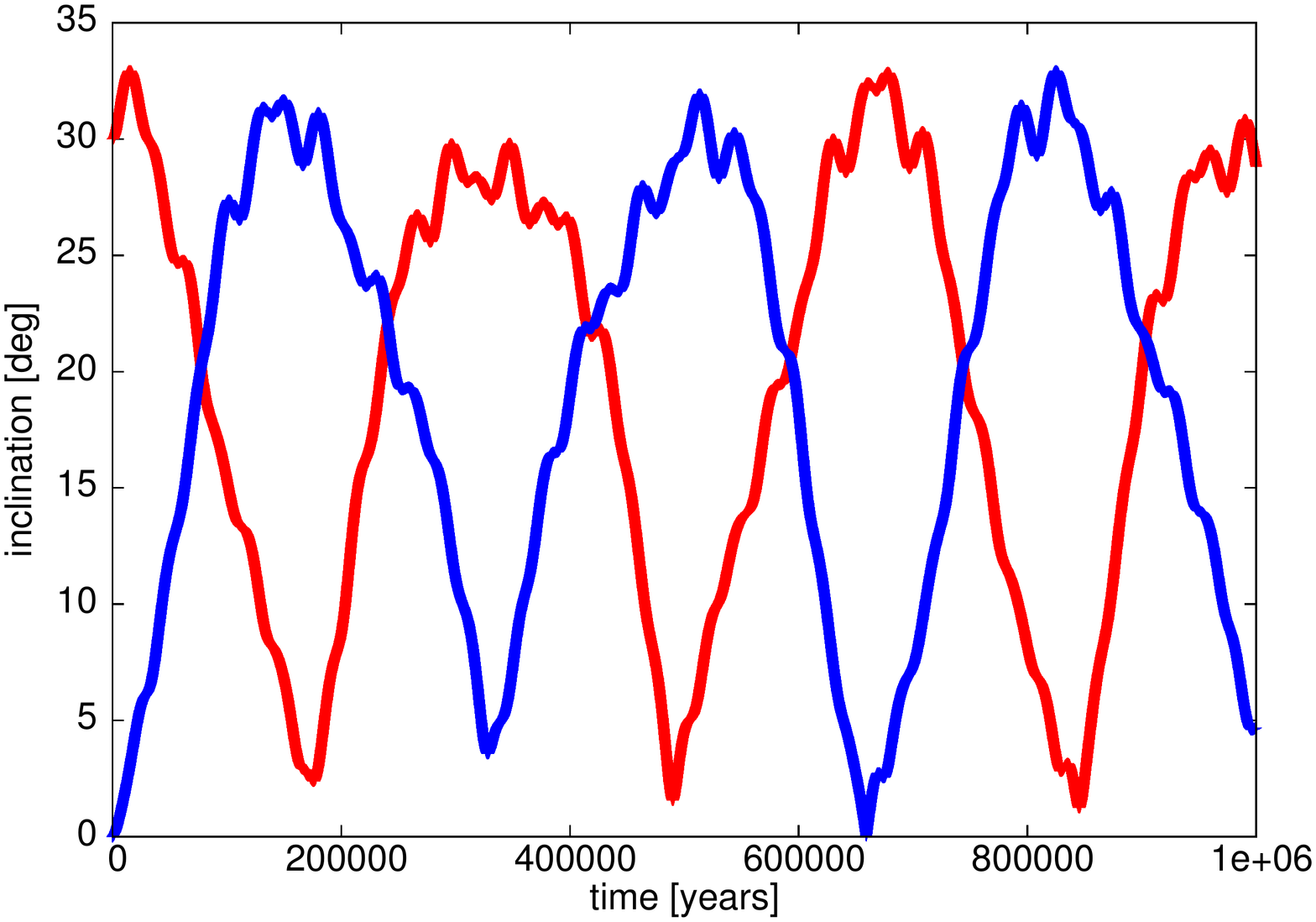}
\includegraphics[width=4.2cm, angle=270]{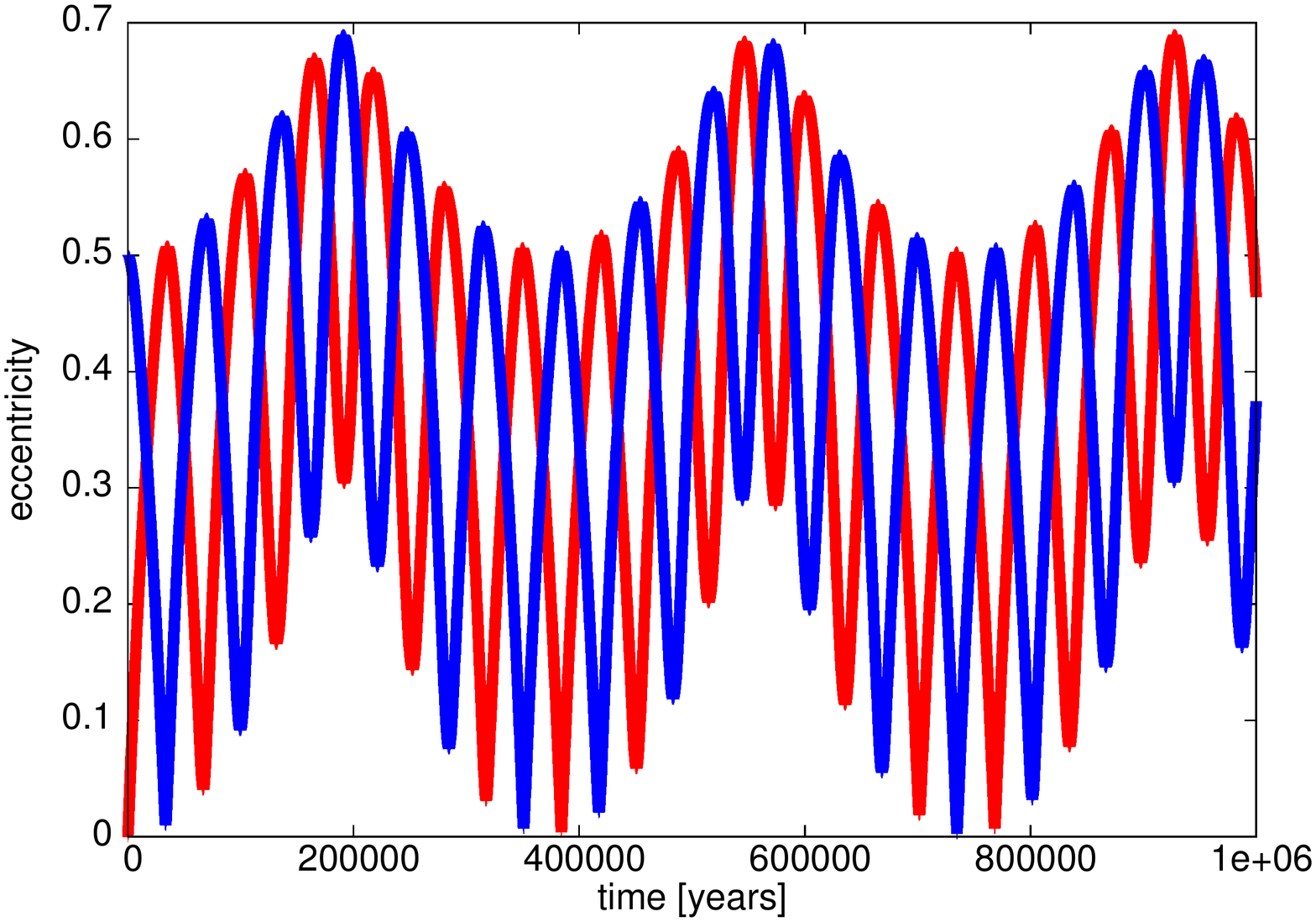}
\includegraphics[width=4.2cm, angle=270]{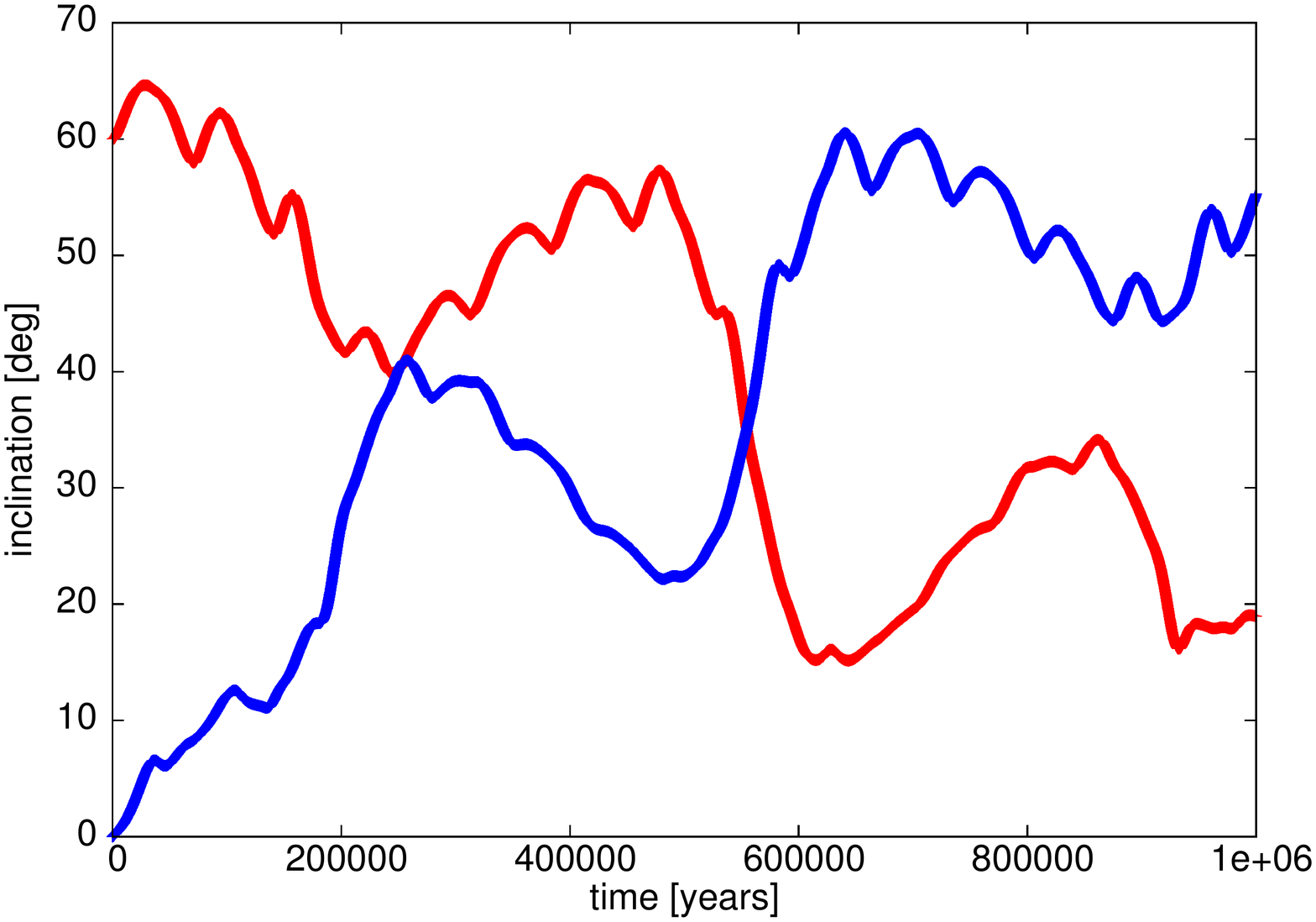}
\caption{Eccentricity (left panel) and inclination (right panel) for three different inclinations (upper graphs: i = $10^{\circ}$, middle graphs: i = $30^{\circ}$, lower graphs: i = $60^{\circ}$). The exchange eccentricity was set to 0.5.}
\label{orbele}
\end{figure}

We performed integrations for $10^{6}$ years in the spatial 3-body problem, consisting of a host star with 1 $M_{Sun}$ and two massive planets with 1 $M_{Earth}$ each, where one of the planets in xch-e motion moves on an inclined orbit. The results are shown in Fig. \ref{unper}, where we plot the exchange eccentricity versus the inclination of P2. The color code corresponds to the $\Delta$a values, where $\Delta a \le 0.001$ was defined as stable xch-e orbit.
From Fig. \ref{unper} one can see that in the planar case stable motion is possible for eccentricities higher than 0.04. For increasing inclination this border is shifted towards higher eccentricities. It turned out that stable xch-e motion is possible for eccentricities higher than $e$ $\approx$ 0.4 up to an inclination of $i$ = $60^{\circ}$. To verify this results we give in Fig. \ref{orbele} the eccentricity and the inclination for three example orbits, where we show the eccentricity (left panel) and the inclination (right panel) for three different inclinations (i = $10^{\circ}$ (first row), $30^{\circ}$ (second row), $60^{\circ}$ (third row)). The exchange eccentricity of the two planets was always set to $e$ = 0.5.\\
As one can see the exchange in eccentricity happens very regularly in all three cases, where for higher inclinations one can see clearly a further overlapping harmonics. In the right panel of Fig. \ref{orbele} one can see the changes in inclination. While for the to lower inclinations (i = $10^{\circ}$ and $30^{\circ}$) the change of inclination happens again very regularly, one can see in the third graph (i = $60^{\circ}$) some irregularities. According to this results we can suppose, that the high inclined orbits will not be long time stable. To clarify the behavior of the high inclined region further studies need to be done.\\
In the second part of this paper, where we investigate the influence of a perturbing body, we therefore used only the planar case (both planets in xch-e motion move in the same plane).

\subsection{The perturbed case}
\label{planar}

\subsubsection{The planar case}

\begin{figure}
\centering
\includegraphics[width=7cm,angle=270]{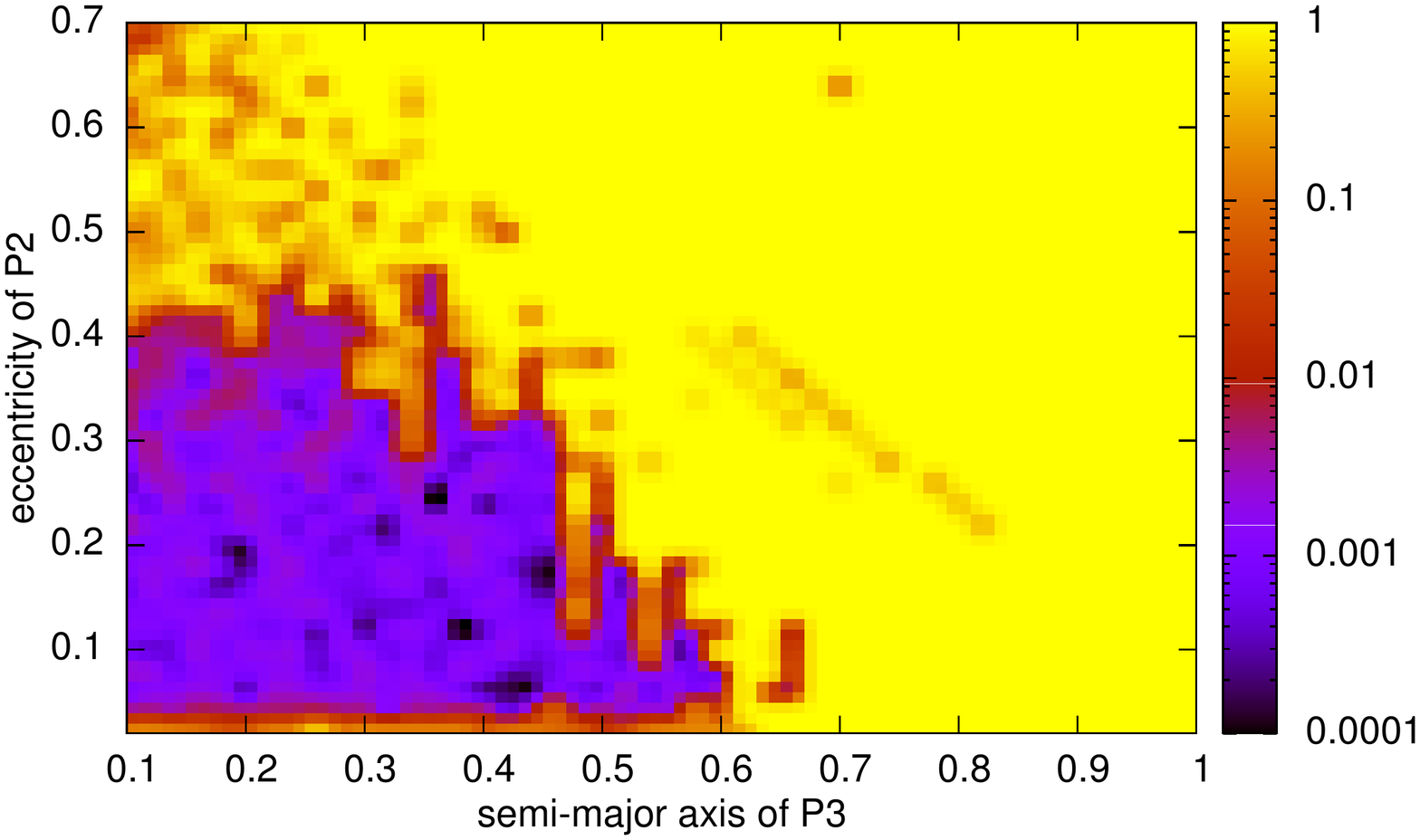}
\caption{Space parameter (a,e) of xch-e orbits, when a perturbing body moves between the star and the two exchange planets. On the x-axis we show the semi-major axis of P3 and on the y-axis the eccentricity of the xch-e orbits. The color code gives the $\Delta a$ values, where black and blue shows stable motion.}
\label{eei_planar}
\end{figure}

In a next step we added a further planet (m = 1 $m_{Jupiter}$) to the system and investigated its influence on the stability of the two planets in xch-e orbit, by changing the semi-major axis and the inclination of the perturber. In Fig. \ref{eei_planar} we show the results, where on the x-axis we show the semi-major axis of P3 and on the y-axis the eccentricity of the xch-e orbits. The color code gives the $\Delta a$ values, where $\Delta a < 0.001$ was defined as stable.
From the figures one can see that no stable xch-e motion is possible, if the semi-major axis of P3 is higher than 0.6 AU and if the exchange eccentricity is lower than 0.06. When P3 moves inside 0.6 AU xch-e motion is possible up to an exchange eccentricity of $e \approx 0.45$. For higher initial exchange eccentricities no stable orbits can be found.\\
In the following section we will investigate the influence of inclined orbits of P3.

\subsubsection{The spatial case}
\label{spatial}
\begin{figure}
\centering
\includegraphics[width=7cm,angle=270]{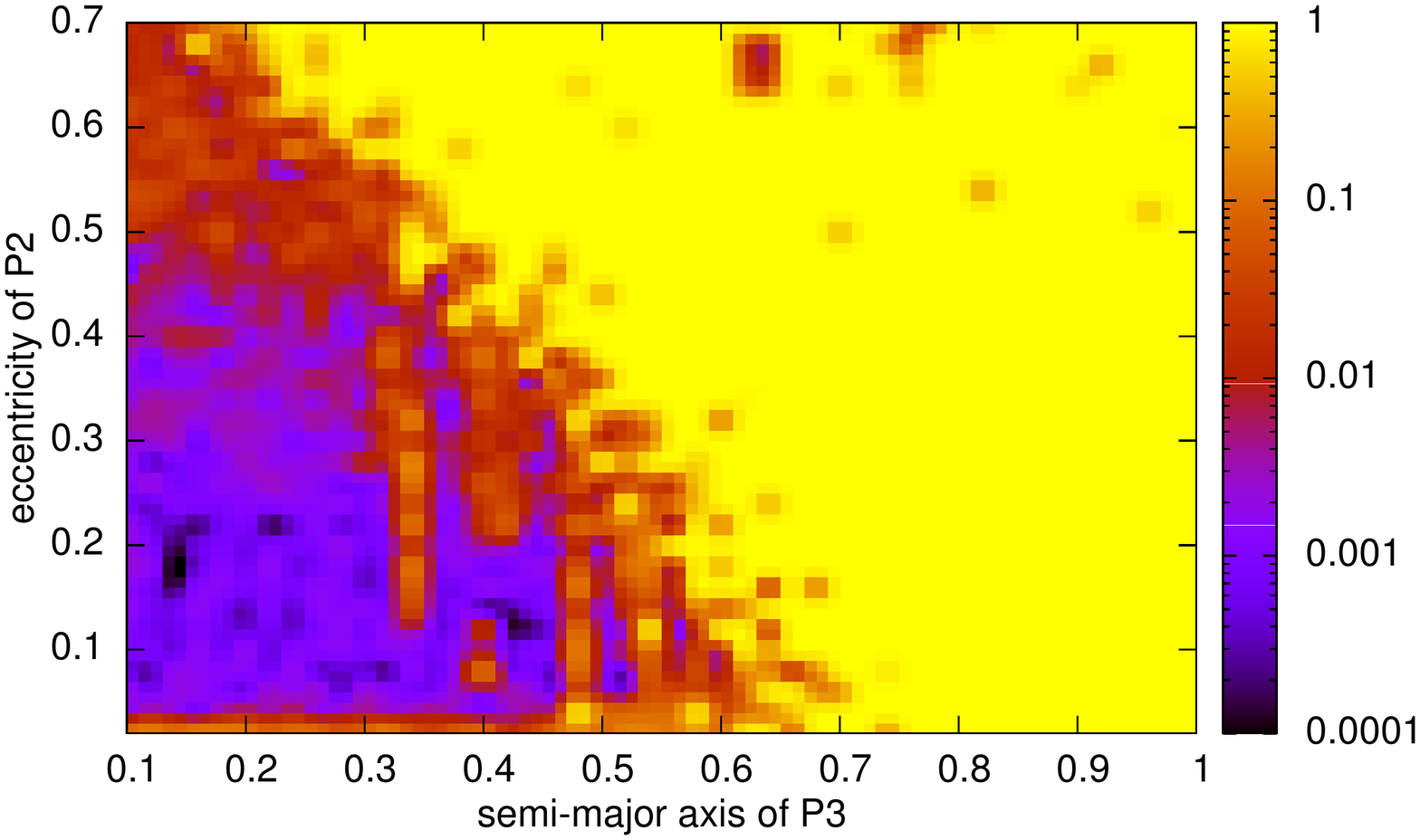}
\caption{ Like Figure \ref{eei_planar}, but for an inclination of the perturbing body of i = $56^{\circ}$.}
\label{eei}
\end{figure}

\begin{figure}
\centering
\includegraphics[width=4.2cm, angle=270]{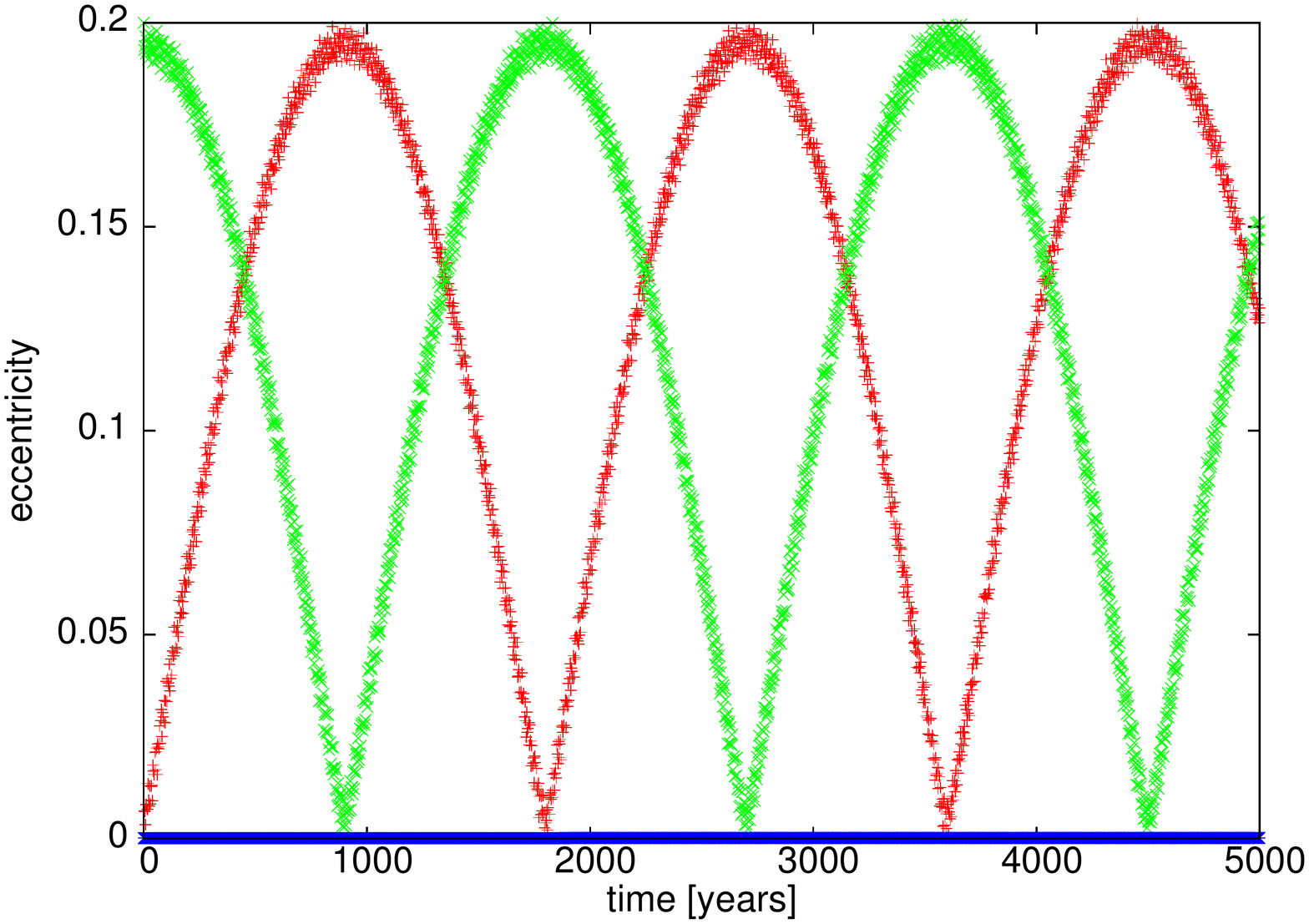}
\includegraphics[width=4.2cm, angle=270]{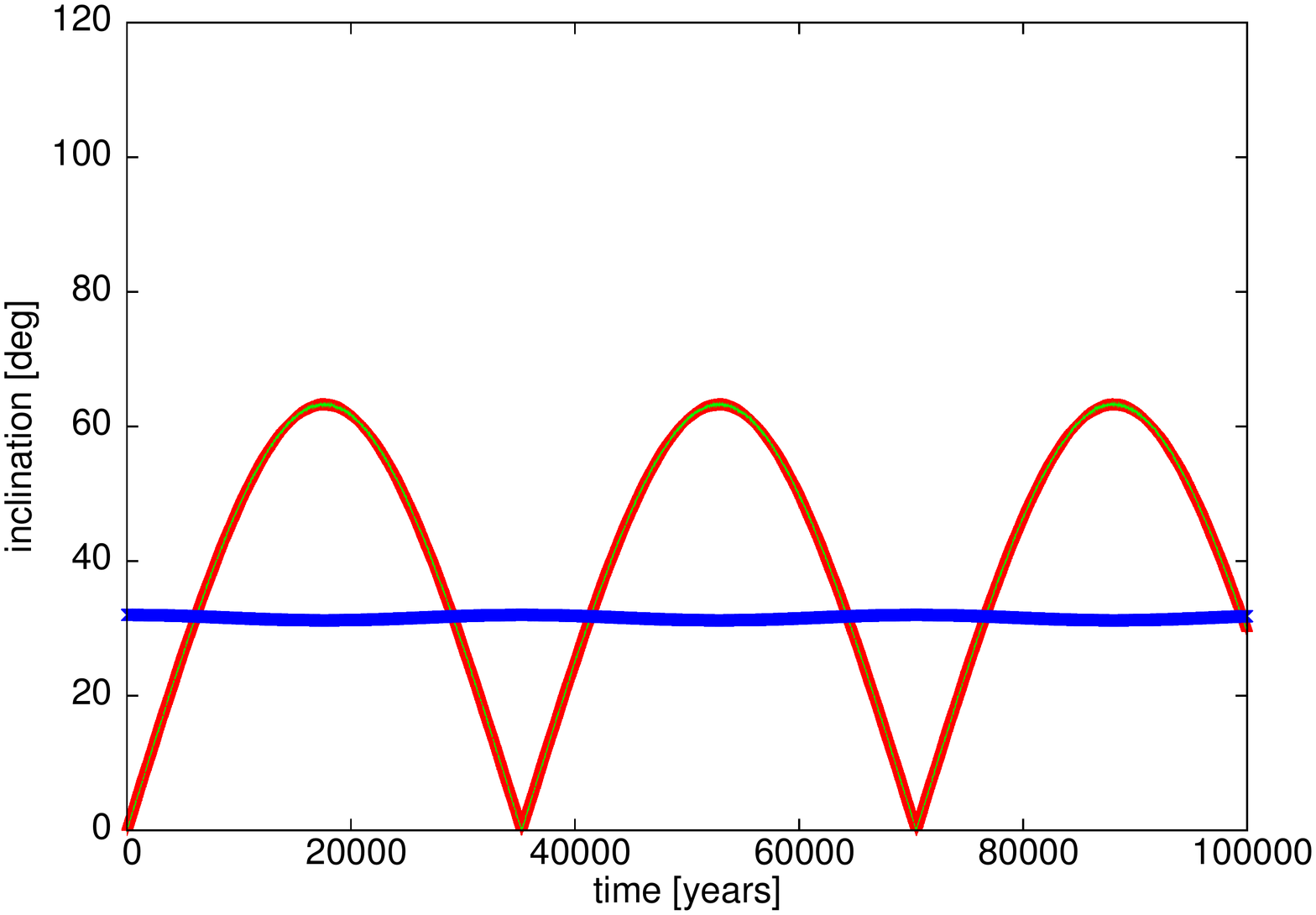}
\includegraphics[width=4.2cm, angle=270]{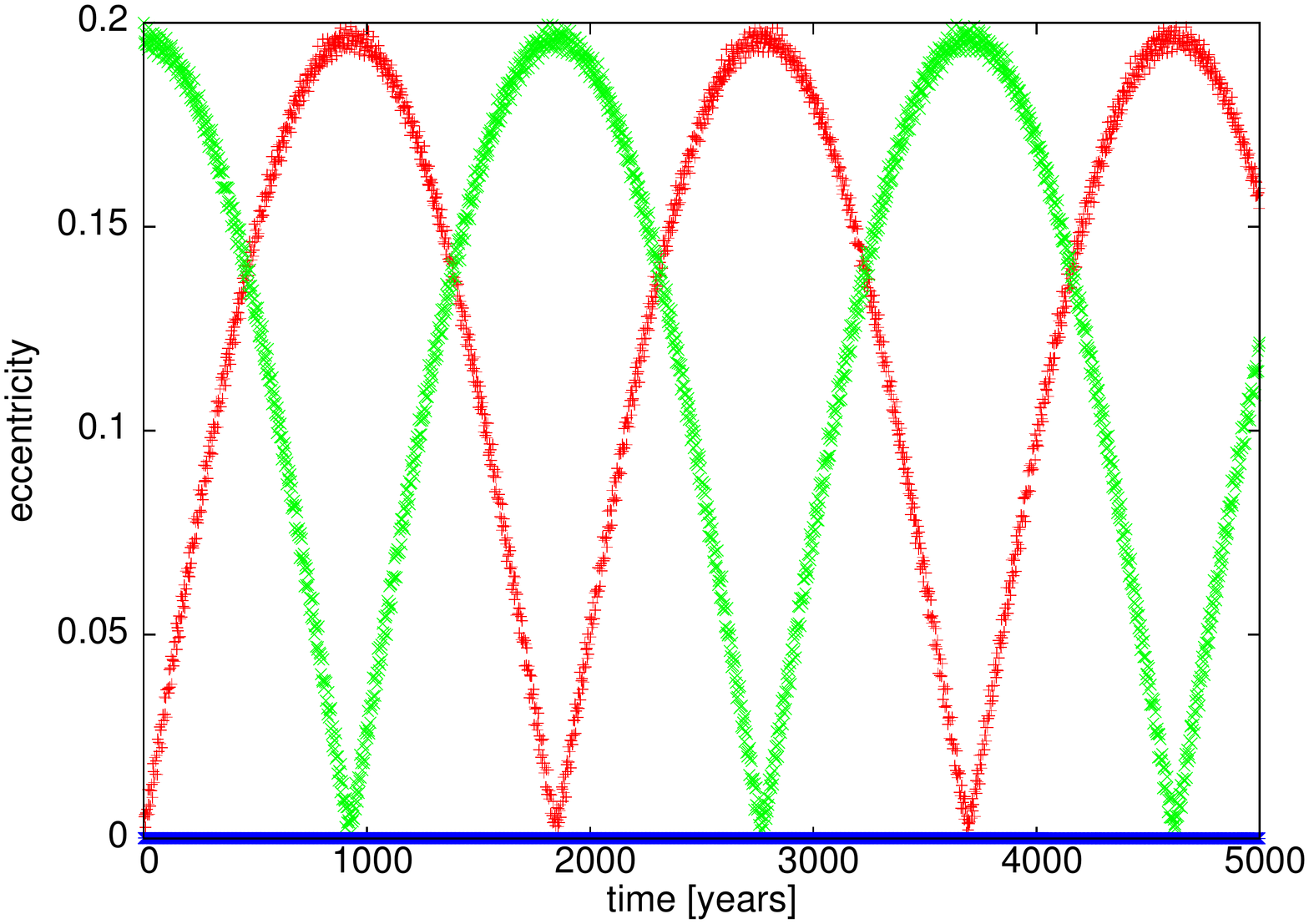}
\includegraphics[width=4.2cm, angle=270]{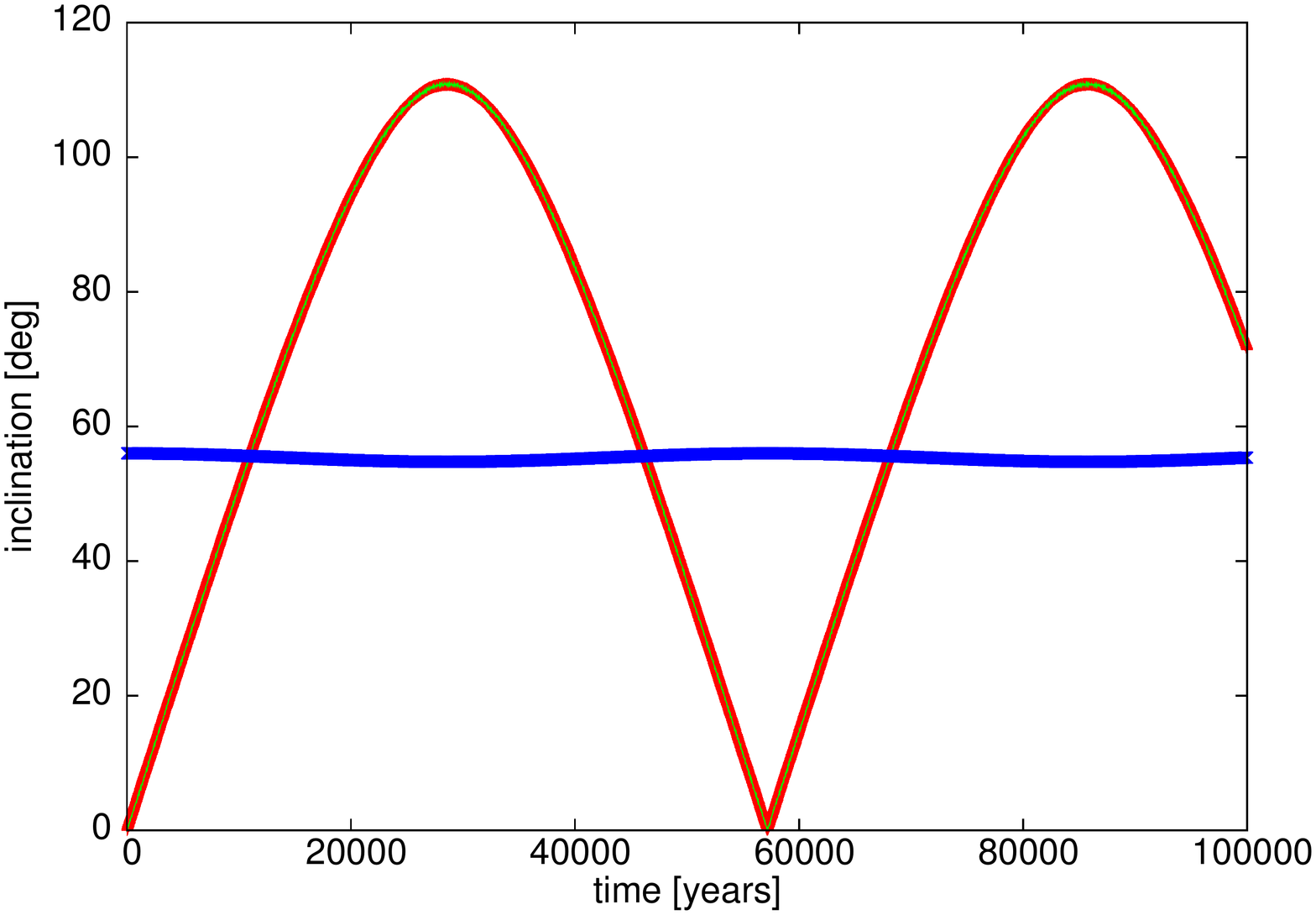}
\caption{Eccentricity (left panel) and inclination (right panel) for three different inclinations of P3 (upper graphs: i = $32^{\circ}$, middle graphs: i = $56^{\circ}$). P3 was always started at 0.2 AU and the eccentricity of the terrestrial planets was 0.2. Color code: Blue: orbital elements of P3; Green and red: orbital elements of the terrestrial planets in xch-e motion.}
\label{elements}
\end{figure}

Here we repeated the calculations for the planar case, but for different inclinations of P3 (i = $8^{\circ}$ - $64^{\circ}$). In Fig. \ref{eei} we show the result for i = $56^{\circ}$, where on the x-axis we give the semi-major axis of P3 and on the y-axis the eccentricity of the xch-e orbits. The color code gives the $\Delta a$ values, where $\Delta a < 0.001$ was defined as stable.\\
Comparing Fig. \ref{eei_planar} with Fig. \ref{eei} one can see that there are just small differences. The stability border shifts slightly from 0.6 AU for i = $0^{\circ}$ and $8^{\circ}$ to 0.45 AU for i = $56^{\circ}$. The highest possible exchange eccentricity remains at approximately 0.4.\\
From our results we can conclude that also a quite high inclined perturber does not destabilize the two planets in xch-e orbits.\\
Since this results are quite unexpected we investigated in a next step in detail the orbital elements of the planets in xch-e motion and of the perturber P3, to check if the terrestrial planets still perform an xch-e orbit. In Fig. \ref{elements} we show the evolution of the eccentricity (left panel) and the inclination (right panel) of the planets in xch-e orbits and the perturber. The calculations were done for a perturber at $a_{P3}$ = 0.2 AU and exchange eccentrixity of $e_{P2}$ = 0.2 for two different inclinations of the perturber ($i_{P3}$ = $32^{\circ}$ (first row), $56^{\circ}$ (second row)).\\
In the first row the inclination of P3 was set to $32^{\circ}$. The two terrestrial planets exchange there eccentricity very nicely, but also the inclination of the planets in exchange orbit become as high as $i_{P1,P2}$ $\approx$ $64^{\circ}$. In the eccentricity and the inclination of P3 one can also see slightly changes and also the xch-e period becomes slightly larger. In the second row the inclination of P3 was set to $56^{\circ}$ and again here the behavior is quite similar. Now the inclination of the terrestrial planets rise up to approximately $115^{\circ}$. While the period of the changes in inclination rises significantly, the xch-e period becomes again just slightly larger.\\
So we can conclude that stable xch-e motion is possible also in the presence of a quite high inclined perturber and the inclination of the xch-e orbits is approximately two times higher than that of the perturber.

\section{Conclusion}

In this investigations we have for the first time studied in detail the two
different types of exchange orbits of planets in the 1:1 MMR. We extended
existing results as we included a possible inclination of the planets P1
and P2 on one hand and on the other hand we took into account perturbations inside or outside
the exchange orbits.\\
For the xch-a orbits it turned out that stable region stays almost constant up to an inclination of $i \approx 8^{\circ}$ and then shrinks with larger inclinations and 
disappears completely for $i > 20^{\circ}$. Additionally we could show that a larger planet P2 increases the stable region. In all four perturbed cases of xch-a motion we could show that the dependence of the largeness of the stable region on the inclination is well represented by a parabola.\\
In the case of xch-e orbits it turned out that stable motion is possible up to an inclination of approximately $i = 60^{\circ}$ of one of the two planets (3-body problem). The investigation of the influence of a perturber showed that stable xch-e motion is possible even in the presence of a quite high inclined perturber, where the inclination of the xch-e orbits becomes approximately two times higher than that of the perturber.\\
It seems that the xch-e orbits may be only of theoretical
interest but as nature has shown us often big surprises (e.g. Janus and
Epimetheus) also this configuration may be found in future. The situation is different for xch-a orbits because we know there exists the
above cited example even in our Solar System. And these orbits resist even
perturbations of other masses in the Saturn system. In extrasolar systems we
have found many 'hot Jupiters' and we could expect that in addition to a
close by massive planet two different sized planets may exist in a larger
distance -- maybe even in a habitable zone --  which sometimes
are opposite to the hosting star and sometimes quite close during their change
of the orbits. Within the last almost 20 years we have discovered so many
different architectures of planetary systems that even two planets on xch-a
orbit may be possible. Let us wait for future discoveries.

\begin{acknowledgements}
BF and RS wants to acknowledge the support by the Austrian FWF project
P23810-N16. RD wants to acknowledge the support by the Austrian FWF NFN project S 11603-N16.
\end{acknowledgements}


\end{document}